\documentclass[reprint,amsmath,amssymb,aps,prx,superscriptaddress,bibnotes,floatfix]{revtex4-2}

\usepackage{tikz} 
\usepackage{graphicx}%
\usepackage{multirow}%
\usepackage{amsmath,amssymb,amsfonts}%
\usepackage{amsthm}%
\usepackage{mathrsfs}%
\usepackage{xcolor}%
\usepackage{textcomp}%
\usepackage{booktabs}%
\usepackage{algorithm}%
\usepackage{algorithmicx}%
\usepackage{algpseudocode}%
\usepackage{listings}%
\usepackage{braket}
\usepackage{tabularx}  
\usepackage{comment}
\usepackage{hyperref}

\usepackage[normalem]{ulem}

\definecolor{vibrant}{HTML}{E77500}
\definecolor{muted}{HTML}{994400}

\hypersetup{
    colorlinks,
    linkcolor = vibrant,
    citecolor = muted,
    urlcolor = muted
}
\begin{document}

\title{Observation of synchronization between two quantum van der Pol oscillators in trapped ions}

\author{Jiarui Liu}\thanks{These authors contributed equally to this work.}
\affiliation{Department of Physics, University of California, Berkeley, CA 94720, USA}
\affiliation{Challenge Institute for Quantum Computation, University of California, Berkeley, California 94720, USA.}
\affiliation{Lawrence Berkeley National Laboratory, Berkeley, CA 94720, USA}

\author{Qiming Wu}\thanks{These authors contributed equally to this work.}
\email[Corresponding author ] {qiming.wu@berkeley.edu}
\affiliation{Department of Physics, University of California, Berkeley, CA 94720, USA}
\affiliation{Challenge Institute for Quantum Computation, University of California, Berkeley, California 94720, USA.}
\affiliation{Lawrence Berkeley National Laboratory, Berkeley, CA 94720, USA}

\author{Joel E. Moore}
\affiliation{Department of Physics, University of California, Berkeley, CA 94720, USA}
\affiliation{Challenge Institute for Quantum Computation, University of California, Berkeley, California 94720, USA.}
\affiliation{Lawrence Berkeley National Laboratory, Berkeley, CA 94720, USA}

\author{Hartmut Haeffner}
\email[Corresponding author ]{hhaeffner@berkeley.edu}
\affiliation{Department of Physics, University of California, Berkeley, CA 94720, USA}
\affiliation{Challenge Institute for Quantum Computation, University of California, Berkeley, California 94720, USA.}
\affiliation{Lawrence Berkeley National Laboratory, Berkeley, CA 94720, USA}

\author{Christopher W. W\"achtler}
\email[Corresponding author ] {cwwaechtler@icmm.csic.es}
\affiliation{Instituto de Ciencia de Materiales de Madrid (ICMM-CSIC), 28049 Madrid, Spain}
\affiliation{Department of Physics, University of California, Berkeley, CA 94720, USA}

\date{\today}

\begin{abstract}
Synchronization is a hallmark of collective behavior that emerges when nonlinear systems interact, spanning scales from mechanical oscillators to planetary orbits.  As a universal phenomenon it underpins the study of complex systems and has far-reaching technological implications. While classical synchronization has a long and rich history, it has not been observed experimentally between multiple quantum limit-cycle oscillators despite a decade of theoretical investigations. We realize synchronization between two quantum van der Pol oscillators by engineering dissipation in a mixed-isotope trapped-ion quantum simulator. The synchronized state is encoded in a fixed relative phase between the oscillators that is inaccessible to local measurements and only revealed through joint readout of both oscillators, in stark contrast to the classical case where synchronization can be observed via individual phase measurements. We further show that the relative phase can be precisely controlled, and that the chain of two oscillators can synchronize to an external field, suggesting applications in sensing. Our results provide a promising pathway for studying more complex synchronized quantum dynamics beyond two oscillators, where a theoretical treatment becomes increasingly challenging, and it remains to be understood whether genuinely quantum features persist in such cases.

\end{abstract}


\maketitle

\section{Introduction}

Self-sustained oscillations emerge when internal feedback mechanisms regulate energy input to compensate for dissipation~\cite{pikovsky1985universal, Juzar2018, Wächtler_2019}. These oscillations correspond to closed and attractive trajectories in phase space, so called limit-cycles. Because of their phase freedom, limit cycle oscillators can entrain to an external drive or mutually synchronize~\cite{laskarObservationQuantumPhase2020b, koppenhoferQuantumSynchronizationIBM2020a, PhysRevResearch.5.033209,liExperimentalRealizationSynchronization2025,vanagOscillatoryClusterPatterns2000}. Among classical models, the van der Pol (vdP) oscillator has served as the canonical example of a limit-cycle system: originally introduced to describe nonlinear electrical circuits in early radio technology~\cite{pikovsky1985universal}, it has since provided the conceptual foundation for synchronization theory across diverse domains, from neuroscience to circadian rhythms~\cite{fitzhughImpulsesPhysiologicalStates1961,jewettRefinementLimitCycle1998}. It is therefore natural that efforts to formulate synchronization in the quantum regime start from the quantum vdP oscillator~\cite{lee2013quantum,  walter2015quantum}. In the quantum regime~\cite{ xuSynchronizationTwoEnsembles2014a, walter2015quantum, rouletQuantumSynchronizationEntanglement2018a, rouletSynchronizingSmallestPossible2018a, schmolkeNoiseInducedQuantumSynchronization2022b, murtadhoCooperationCompetitionSynchronous2023b, laskarObservationQuantumPhase2020b, koppenhoferQuantumSynchronizationIBM2020a, PhysRevResearch.5.033209,liExperimentalRealizationSynchronization2025, PhysRevA.105.062206, PhysRevLett.120.163601}, a limit cycle is defined as an attractive solution of the dissipative dynamics with a steady state exhibiting $U(1)$-symmetry. The simplest realization of such a quantum limit cycle is a harmonic oscillator with incoherent one-phonon gain and two-phonon loss processes, which together define the quantum vdP oscillator~\cite{lee2013quantum}.


The quantized motional modes of trapped ions provide an excellent platform for realizing quantum limit-cycle oscillators and  exploring their synchronized dynamics. Their long coherence times, precise controllability, and strong coupling to internal states~\cite{broz2023test,macdonell2023predicting,wu2023continuous} accommodate the implementation of quantum dynamics in bosonic degrees of freedom~\cite{gottesman2001encoding,braunstein2005quantum}, which has enabled a wide range of applications  including analog quantum simulation, quantum-enhanced metrology, and fault-tolerant quantum computing with bosonic codes~\cite{whitlow2023quantum,gorman2018engineering,valahu2023direct,burd2019quantum,mccormick2019quantum,fluhmann2019encoding,matsos2024universal}. However, the lack of direct observables for the motional states has traditionally limited the exploitation of quantum resources in the infinite-dimensional Hilbert space. Recent proposals and experimental realizations have introduced efficient benchmarking techniques for motional states~\cite{cimini2020neural,fluhmann2020direct,valahu2024benchmarking}. These advancements, combined with programmable dissipation and noise injection~\cite{behrle2023phonon,so2024trapped}, enable the level of control and measurement required for synchronization experiments.

\begin{figure*}
    \centering
    \includegraphics[width=0.9\linewidth]{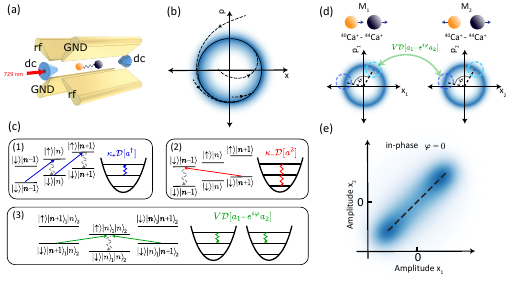}
    
\caption{Synchronization of quantum van der Pol oscillators with trapped ions.
(a) Schematics of experimental setup with a $\mathrm{^{40}Ca^+}-\mathrm{^{44}Ca^+}$ ion crystal trapped in a linear rf Paul trap. The two axial motional modes are used for the synchronization experiment, where the state of the $\mathrm{^{40}Ca^+}$-ion and the motional state of the ion crystal are controlled by a 729\,nm laser light.
(b) Illustration of the donut-shaped Wigner function of a quantum limit-cycle oscillator. Black dashed curves indicate classical trajectories attracted towards the classical limit cycle.
(c) Experimental implementation of the corresponding dissipator to generate and synchronize two vdP oscillators. (1) Negative damping is realized by driving the blue sideband of $M_i$, followed by a qubit reset; (2) Nonlinear damping is realized by driving the second order red sideband of $M_i$, followed by a qubit reset; (3) Collective dissipation is generated by simultaneous driving the red sidebands of $M_1, M_2$ with a phase difference $\varphi$, followed by another qubit reset. (d) 
The Wigner function of each mode $M_1$ and $M_2$ individually does not show any phase preference. Instead, the synchronized dynamics only appears as stable relative phase relation between the two oscillators, which can only be observed by a joint measurement. 
(e) Joint probability distribution $P(x_1,x_2)$ for in-phase ($\varphi=0$) synchronized  oscillators. Black dashed line shows the corresponding classical trajectory.}

    \label{fig1_exp_overview}
\end{figure*}

Here, we experimentally demonstrate synchronization between two quantum vdP oscillators encoded in the motional state of two motional modes of a two
ion crystal. 
The vdP oscillators are generated through a combination of two-phonon loss and one-phonon gain processes of the motional modes utilizing an auxiliary qubit. We show that 
the steady states of the vdP oscillators are highly tunable from the near-classical to the quantum regime via the gain–loss ratio, and the resulting limit-cycle behavior is revealed by tracking the motional-state dynamics via Wigner function reconstruction. We achieve synchronization via engineered dissipative coupling~\cite{lee2014entanglement}, confirmed by reconstructing the joint probability distribution of the two-mode quantum states and supported by numerical simulations. We further show that synchronization is robust against detuning between the effective frequencies of the oscillators. Finally, we probe phase diffusion during mutual synchronization by phase-locking one oscillator to an external drive and measuring the Wigner functions of both modes at various detunings. Our results pave the way for investigating synchronized dynamics in networks of quantum limit cycles~\cite{matheny2019exotic}, harnessing engineered dissipation as a resource for complex quantum dynamics, and developing new applications in quantum metrology.

\begin{figure*}[hbt!]
    \centering
    \includegraphics[width=1.0\linewidth]{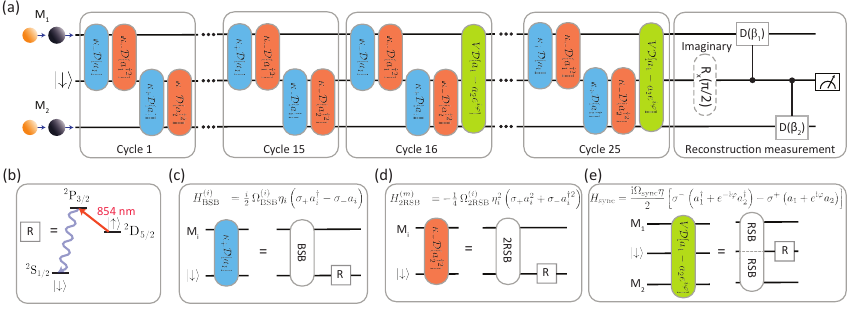}
    \caption{The quantum circuit for synchronization between two quantum vdP oscillators and reconstruction of the motional states. (a) The circuit acts on one qubit and two motional modes, where the auxiliary qubit is to generate collective dissipation on the two vdP oscillators and to read out the two-mode state. Each cycle consists of a stroboscopic application of negative and nonlinear dampings on both modes [described in (c)(d)]. We repeat the sequence fifteen times to generate the two vdP oscillator states, with another ten cycles including the dissipative coupling [described in (e)] to synchronize them. For the motional reconstruction, the controlled displacement in the $\sigma_x$ basis (realized by SDF) maps the motional information onto the qubit. The imaginary part is read out with an additional $\pi/2$ qubit rotation along the $x$ direction prior to state detection. (b) The qubit reset is realized with  854\,nm repumping laser light to reset it to the $\ket{\downarrow}_z$ state. (c)(d)\,The negative (nonlinear) damping on $M_i$ is realized by a coherent BSB (2RSB) drive followed by a qubit reset R. (e) Collective dissipation on the two motional modes is realized by simultaneously driving the RSB on $M_1$ and $M_2$ with a constant phase difference $\varphi$, followed by a qubit reset [see details in Appendix~\ref{app:Experimental_platform}] .}
    \label{fig_ex:quantum_circuit}
\end{figure*}

\section{Synchronizing quantum van der Pol oscillators}



\subsection{Quantum vdP oscillators in trapped ions}

To experimentally realize two coupled vdP oscillators we employ the two axial motional modes, $M_1$ (in-phase) and $M_2$ (out-of-phase), of a $\mathrm{^{40}Ca^+}$--$\mathrm{^{44}Ca^+}$ linear two-ion crystal confined in an rf Paul trap [see Fig.~\ref{fig1_exp_overview}(a)]. The uncoupled collective motion is described by the Hamiltonian $ H_0=  \omega_1 a_1^\dagger  a_1+ \omega_2 a_2^\dagger a_2$, where $\omega_{1,2}$ are the frequencies of mode $M_{1,2}$ and $a_{1,2}$ are the corresponding ladder operators. Qubit-motion couplings are implemented via driving motional sidebands of the $4^2 S_{1/2}\leftrightarrow 3^2 D_{5/2}$ quadrupole transition of the $\mathrm{^{40}Ca^+}$ ion. Due to the $\mathrm{^{40}Ca^+}$-
$\mathrm{^{44}Ca^+}$ isotope shift of 5.34\,GHz on this transition~\cite{knollmann2019part}, the applied laser field does not couple to the internal state of the $\mathrm{^{44}Ca^+}$ ion, resulting in Jaynes-Cummings-type interactions. Motional dissipation required for the limit-cycle dynamics of the quantum vdP oscillators is engineered by returning the $3^2 D_{5/2}(m=-5/2)$ population to the $4^2 S_{1/2}(m=-1/2)$ state via optical pumping, i.e. resetting the qubit \{$\ket{\uparrow}=D_{5/2}(m=-5/2),\ket{ \downarrow}=S_{1/2}(m=-1/2)$\}  [see Fig.~\ref{fig1_exp_overview}(c) and Figs.~\ref{fig_ex:quantum_circuit}(b)-(d)]. Specifically, negative damping ($\mathcal{D}[a^\dagger]$) is implemented through coupling to the first-order blue sideband (BSB), which corresponds to transitions that add one phonon. Nonlinear damping ($\mathcal{D}[a^2]$) is realized by driving the second-order red sideband (2RSB), which corresponds to transitions that remove two phonons simultaneously. In both cases, the interaction is applied stroboscopically with a randomized phase and followed by resetting the qubit after each evolution step, ensuring effective damping in the individual motional modes.

To synchronize the two oscillators, we interleave layers of dissipative coupling ($\mathcal{D}[a_1-a_2e^{\rm{i}\varphi}]$) with resonantly driving the first-order red sidebands (RSBs) on the two modes simultaneously with a fixed phase difference $\varphi$, followed by qubit reset [see Fig.~\ref{fig_ex:quantum_circuit}(e)]. Here the auxiliary qubit is used to implement collective dissipation between the modes. The full sequence of individual negative damping, nonlinear damping and dissipative coupling pulses is repeated for $n$ cycles until the system reaches a steady state. The detailed quantum circuit is shown in  Fig.~\ref{fig_ex:quantum_circuit}(a).  The resulting effective dynamics is described by the Lindblad master equation
\begin{equation}
\label{eq:EffectiveModel}
\begin{split}
\dot{\varrho} = &  -\mathrm{i}[\tilde{H}_0, \varrho] 
+ \kappa_+ \sum_{i=1}^2 \mathcal{D}[a_i^\dagger] \varrho 
+ \kappa_- \sum_{i=1}^2 \mathcal{D}[a_i^2] \varrho \\ &
+ V \mathcal{D}[a_1 - a_2 e^{\mathrm{i}\varphi}] \varrho.
\end{split}
\end{equation}
Here, $\tilde{H_0} = \delta_1 a_1^\dagger a_1 +\delta_2 a_2^\dagger b_2 $ with  $\delta_{1,2}$ being the differences  between motional frequencies $\omega_{1,2}$ and half the laser frequency splittings of the BSB and RSB drives,  which set the rotating frame of the oscillators. For convenience we set $\hbar \equiv 1$. The Lindblad dissipators $\mathcal{D}[L] \varrho=L \varrho L^{\dagger}-\left\{L^{\dagger} L, \varrho\right\}/ 2$ describe negative damping with rate $\kappa_+$, nonlinear damping with rate $\kappa_-$, and collective dissipation with strength $V$. Note that the ratio $\kappa_- / \kappa_+$ determines the average phonon number and thus the radius of the donut-shaped Wigner functions.  Moreover, the phase $\varphi$ appearing in Eq.~(\ref{eq:EffectiveModel}) controls the  phase of the dissipative coupling between the oscillators as illustrated in Fig.~\ref{fig1_exp_overview}(c). The details of the experimental protocols can be found in Appendix~\ref{app:Experimental_platform}.

\begin{figure*}[hbt!]
    \centering
    \includegraphics[width=1.0\linewidth]{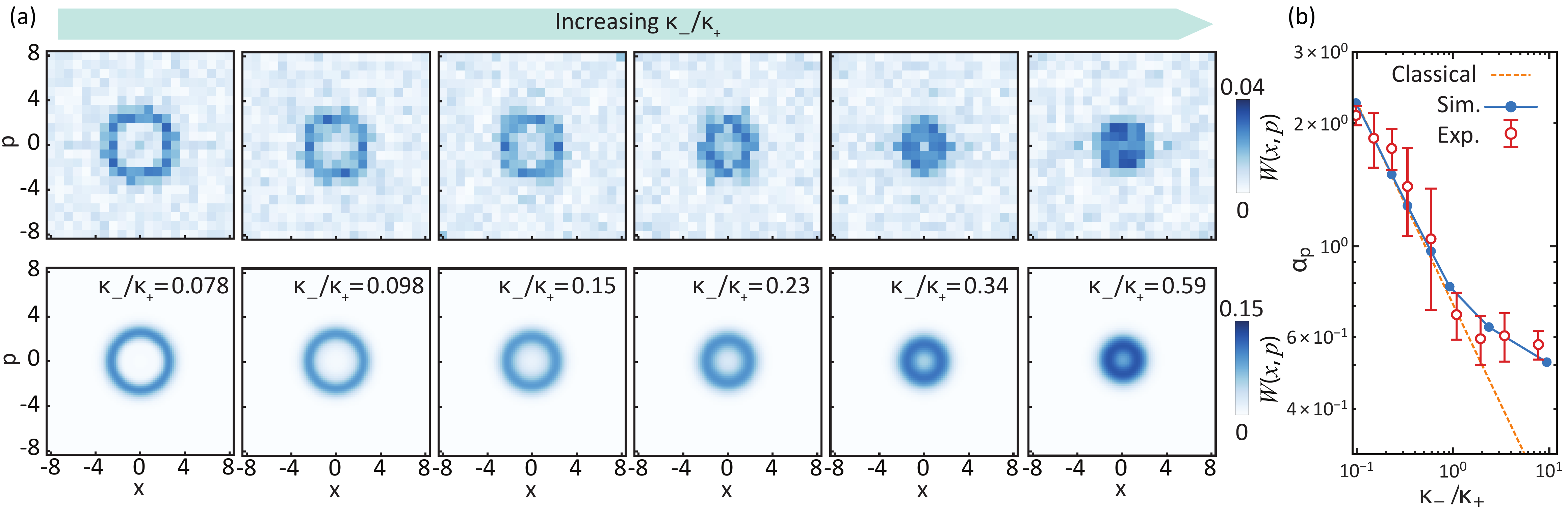}
    \caption{Crossover of the vdP oscillator from the near-classical to the deep quantum regime. (a) Top row shows various experimentally reconstructed Wigner function of vdP oscillator states with different mean phonon number when increasing $\kappa_-/\kappa_+$. The bottom row shows the corresponding numerical simulation. (b) The radius $\alpha_p$ is shown as a function of the ratio 
$\kappa_{-}/\kappa_{+}$ on a logarithmic scale. Red dots are the experimental data, blue solid curve is the numerical simulation using Lindblad equation and the orange dashed line represents the classical mean field theory. When $\kappa_-/\kappa_+$ increases, the oscillator becomes more quantum and deviates from the  classical mean-field theory. During the experiment we keep  $\kappa_+/2\pi=0.112 $\,kHz to be constant and change the strength of $\kappa_-$ by tunning the power of the BSB drive.}
    \label{fig_ex:classical_to_quantum}
\end{figure*}

\subsection{Experimental characterization of a single vdP oscillator}

Before exploring synchronization, we first establish the  control of a single quantum vdP oscillator using  a single $\mathrm{^{40}Ca^+}$ ion.   First, we show that the steady state of the vdP oscillator can be precisely controlled by changing the relative strengths ($\kappa_+$ and $\kappa_-$), which affects the mean phonon number and the radius of the donut-shaped Wigner function. Previous research has classified different parameter regimes of a single vdP oscillator, i.e. the near-classical regime with $\kappa_-/\kappa_+\approx0.1$, the quantum regime with $\kappa_-/\kappa_+\approx 1$ and the deep quantum regime with $\kappa_-/\kappa_+\approx 10$ \cite{mokSynchronizationBoostSinglephoton2020a}.  In Fig.~\ref{fig_ex:classical_to_quantum}(a), we show the experimental reconstructed Wigner function of various vdP oscillator states from near-classical to the deep quantum regime. When the negative damping $\kappa_+$ exceeds $\kappa_-$ and the mean phonon number is large, classical mean field theory predicts  a limit cycle with a radius $\alpha_p=\sqrt{(x^2+p^2)/2}=\sqrt{\kappa_+/2\kappa_-}$ in phase space, where $\alpha_p$ is defined as the most probable position of the vdP Wigner function 
(i.e., the radius of the donut). When the nonlinear damping $\kappa_-$ exceeds $\kappa_+$, the oscillator approaches the ground state upon increasing $\kappa_-$ and $\alpha_p$ starts to deviate from the classical trajectory as shown in Fig.~\ref{fig_ex:classical_to_quantum}(b).

\begin{figure*}[hbt!]
    \centering
    \includegraphics[width=0.85\linewidth]{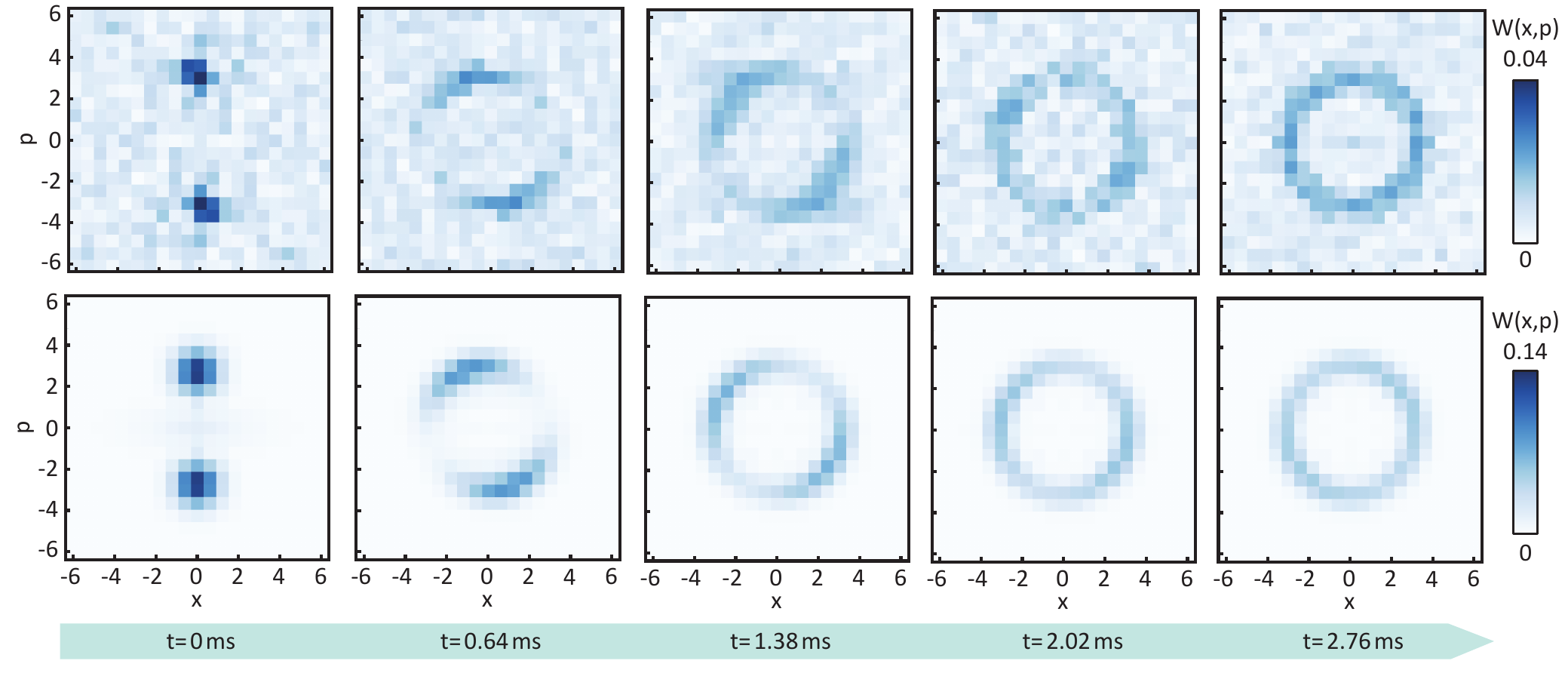}
    \caption{Evolution towards the limit cycle of a single vdP oscillator. We create a cat state with $\alpha=3$ at $t=0$\,ms and measure the time evolution to the vdP oscillator state with Wigner function reconstruction. Top row shows the experimental data and the bottom row shows the numerical simulation. The parameters used in experiment are $t_{\rm 2RSB}=48\,\rm \mu s, t_{\rm BSB}=24\,\rm \mu s$, $\Omega_{\rm BSB}/2\pi=0.06\rm$\, MHz, $\Omega_{\rm 2RSB}/2\pi=0.1$\,MHz. 
    }
    \label{fig_ex:limit-cycle}
\end{figure*}

 Second, we probe that the limit-cycle is in fact the nonequilibrium steady-state (and thus attractive) solution of the dissipative dynamics, which is reached independently of initial conditions. To this end, we measure  the time evolution of Wigner function from a mixture of opposite coherent states ($\varrho=\frac{1}{2}(\ket{\alpha}\bra{\alpha}+\ket{-\alpha}\bra{-\alpha})$)  to the vdP oscillator state. In Fig.~\ref{fig_ex:limit-cycle}, we begin with initial state of a mixture of two opposite coherent state at $t=0$. After applying the vdP interaction, the state gradually diffuses into a donut shape with the steady state radius determined by the ratio of interaction strength $\kappa_-/\kappa_+$.

\begin{figure}[!h]
    \centering
    \includegraphics[width = \linewidth]{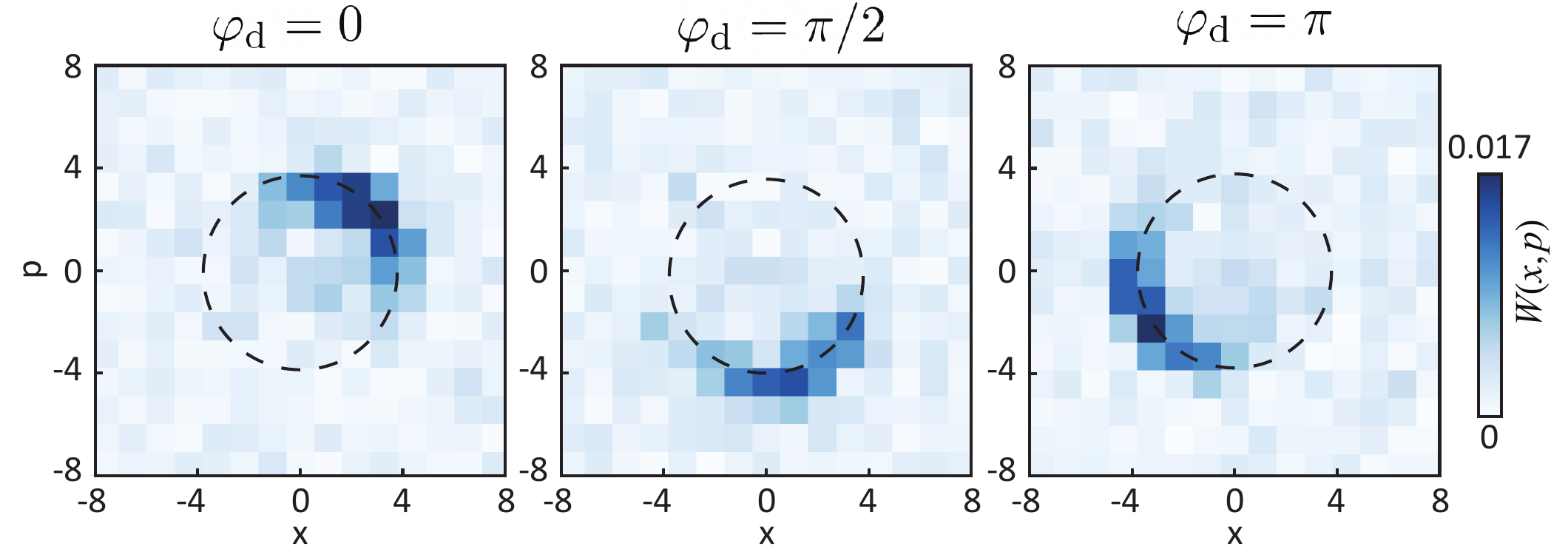}
    \caption{A single vdP oscillator synchronized to an external drive with different  phases $\varphi_d = 0,\,\pi/2,\,\pi$. The dashed circles indicate the size (radius $\alpha_p$) of the vdP oscillator with no external drive.}
    \label{fig_ex:external_drive}
\end{figure}

Third, we show that the phase of a single vdP oscillator can be entrained to an external drive. An isolated vdP oscillator exhibits no intrinsic phase preference. When an external coherent drive $H=\Omega_{d}(a e^{i\varphi_{\rm d}}+a^\dagger e^{-i\varphi_{\rm d}})$ of sufficient strength is introduced, this rotational symmetry is broken and the oscillator’s phase becomes entrained to that of the external drive~\cite{lee2013quantum, walter2014quantum,dutta2019critical, liExperimentalRealizationSynchronization2025}. In practical experimental conditions, the phase of the external drive has a constant offset from the motional phase of the SDF $\phi_m$ for read-out due to differences in the optical and electrical signal paths. In Fig.~\ref{fig_ex:external_drive}, we show the experimental results of a single vdP oscillator locked to the external drive with different phase preferences $\varphi_d = 0,\,\pi/2,\,\pi$. These results also serve as a reference for subsequent experiments on mutual synchronization under the drive (see Sec.~\ref{sec:ExternalDrive}).

\subsection{Mutual synchronization in the near-classical and quantum regime} 
\label{sec:MutualSync}
\begin{figure*}[t!]
    \centering
    \includegraphics[width=1\linewidth]{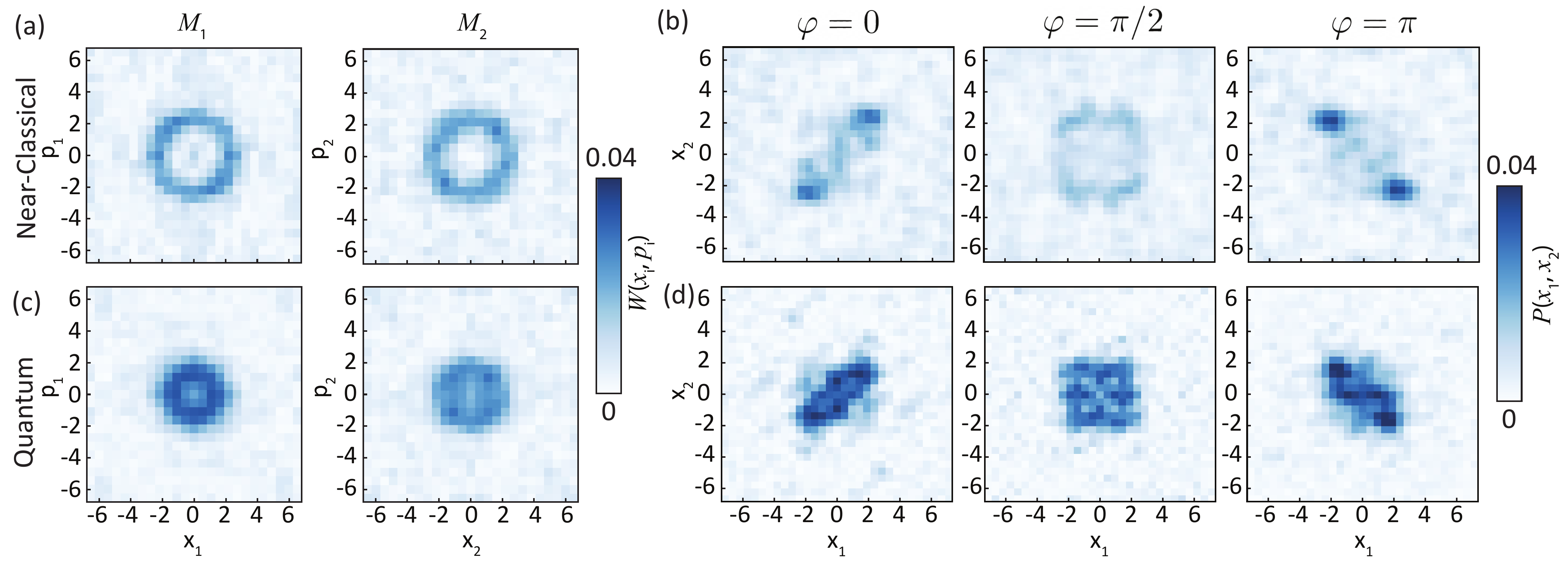}
    \caption{Synchronized dynamics in the near-classical and quantum regime. 
(a)(b) Near-classical regime with $\kappa_- / \kappa_+ = 0.14$ and mean phonon number $\bar{n} = 3.5$. 
(a) Individually 
reconstructed Wigner functions of the two quantum vdP oscillators  show the expected donut-shaped profile without a preferred phase.
(b) The reconstructed joint probability distributions $P(x_1, x_2)$ for different synchronization phases $\varphi$. For $\varphi = 0$ and $\pi$, the two oscillators exhibit in-phase and anti-phase synchronization, indicated by diagonal and anti-diagonal features in the $x_1$--$x_2$ plane. For $\varphi = \pi/2$, the distribution forms a circle, reflecting the relative phase lag between the oscillators.
(c)(d) Quantum regime with $\kappa_- / \kappa_+ = 0.42$ and $\bar{n} = 1.4$.
(c) Reconstructed Wigner functions show reduced limit-cycle amplitudes as the system approaches the ground state.
(d) Despite the increased role of quantum noise, correlations in the $x_1$--$x_2$ plane remain visible, indicating persistence of synchronization in the quantum regime. Wigner functions in (a)(c) are measured at $\varphi = \pi$. Corresponding numerical simulations are provided in Appendix~\ref{app:Experimental_parameters}, showing good agreement with experimental values.}
    \label{fig2}
\end{figure*}


After characterizing the dynamics of a single vdP oscillator, we extend our study to the synchronization between two quantum vdP oscillators. 
To investigate quantum limit-cycle synchronization, we realize two quantum vdP oscillators and synchronize them via collective dissipation.  We first implement it in the near-classical regime by tuning the ratio between the two damping processes to be $\kappa_- / \kappa_+ = 0.14$~\cite{mokSynchronizationBoostSinglephoton2020a}, identical for each oscillator, resulting in a steady-state mean phonon number of $\bar{n}= 3.5$. The effective rates are implemented by adjusting the strengths and durations of the BSB and 2RSB interactions on each mode. The collective dissipative coupling has an effective rate $V_0/2\pi =  0.1$\,kHz (see Appendix~\ref{app:Experimental_parameters}). During the first fifteen cycles, the dissipative coupling is not applied such that the two vdP oscillators reach their steady states independently and without phase preference. Synchronization is then introduced by applying ten additional cycles including the dissipative coupling with a fixed phase $\varphi$. By varying $\varphi$, we gain full control over the relative phase alignment between the two oscillators [see Fig.~\ref{fig2}(b)].

After preparing the desired oscillator states, we perform a readout of the Wigner function for each individual mode. Fig.~\ref{fig2}(a) shows the reconstructed Wigner functions of each mode individually. Both oscillators exhibit similar phonon occupations, resulting in the donut-shaped Wigner function, which is characteristic of a quantum limit-cycle. Because synchronization in our system occurs only in the relative phase, individual oscillators show no signatures of phase locking. This contrasts with externally driven quantum vdP oscillators, where synchronization remains locally observed through single-mode measurements~\cite{liExperimentalRealizationSynchronization2025}. Instead, the global phase of each oscillator remains unpolarized, leading to isotropic Wigner functions. This shows that synchronization is not accessible through local measurements alone, since the phase of each oscillator cannot be individually resolved [see Figs.~\ref{fig1_exp_overview}(d) and (e)]. In classical systems, by contrast, the phase of each oscillator can be measured separately, and synchronization can then be observed by comparing these phases. For the quantum oscillators, however, the relative phase is not locally accessible and only revealed through joint measurements.

\begin{figure*}[ht]
    \centering
    \includegraphics[width=0.9\linewidth]{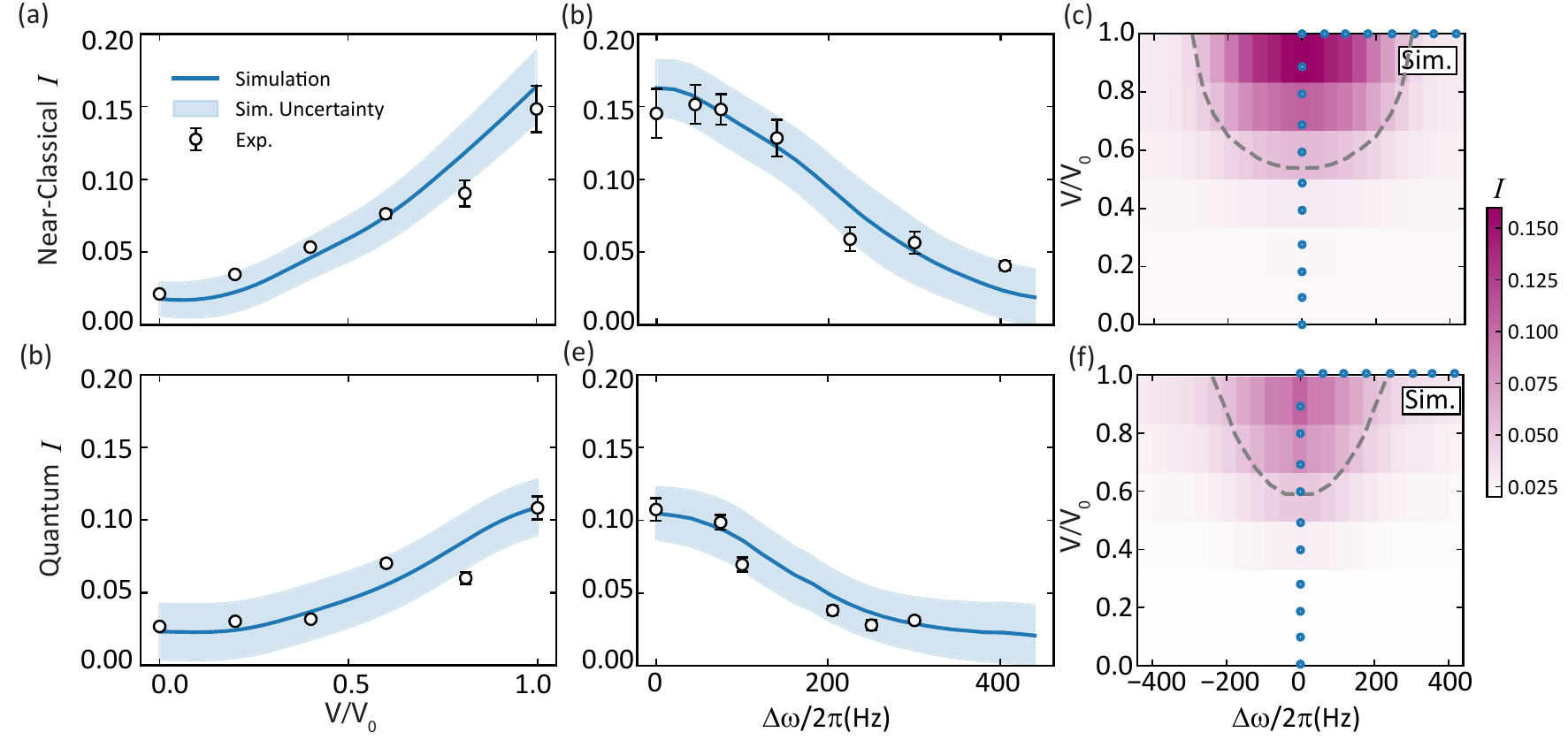}
    \caption{Arnold tongue–like synchronization landscape.
(a)-(c) Mutual information $\mathcal{I}$ as an indicator of synchronization between two quantum vdP oscillators in the near-classical regime. (a) At zero detuning, increasing the dissipative coupling strength enhances synchronization, reflected in higher mutual information. 
(b) At fixed coupling strength, increasing the detuning suppresses synchronization and thus decreases mutual information. $V_0$ is the coupling strength used in Fig.~\ref{fig2}.
(c) Simulation across the full coupling–detuning parameter space reveals the characteristic Arnold tongue–like structure. The gray dashed curve represents the contour for $\mathcal{I}=0.05$.  The blue dots represent the two cuts where the experimental data is taken. The simulations include a global offset accounting for read-out noise.
(d)-(f) Same as (a)-(c), but in the quantum regime. Here, quantum noise reduces the overall level of synchronization, leading to smaller values of mutual information, though the qualitative behavior persists. Black circles in panels (a)(b) and (d)(e) show the experimentally measured $\mathcal{I}$ with error bars reflecting quantum projection noise. Solid lines represent numerical simulations using experimental parameters, and the shaded regions indicate uncertainty from motional frequency drifts and Rabi frequency fluctuations~(see Appendix~\ref{app:Error_analysis}).
}
    \label{fig:Arnoldtongue}
\end{figure*}

To uncover the correlations between the two modes, both oscillators need to be coherently displaced simultaneously, which allows us to access the two-mode characteristic function~\cite{valahu2023direct}
\begin{equation}\label{eq_readout}
    \chi\left(\beta_1, \beta_2\right)=\langle\Psi| \mathcal{D}_1\left(\beta_1\right) \mathcal{D}_2\left( \beta_2\right)|\Psi\rangle,
\end{equation}
where $\mathcal{D}_i(\beta_i)=\exp(\beta_ia_i^\dagger-\beta_i^\ast a_i)$ denotes the coherent displacement on $M_i$, and $\ket{\Psi}$ is the two-mode state. The joint probability distribution for measuring the two modes along general quadrature axes $x_{\phi_j} = (a_j e^{-\mathrm{i}\phi_j} + a_j^\dagger e^{\mathrm{i}\phi_j}) / \sqrt{2}$ is then obtained through the Fourier transform of the characteristic function:
\begin{equation}
\begin{split}
P(x_{\phi_1}, x_{\phi_2}) = &\iint \frac{d\beta_1\, d\beta_2}{2\pi^2} \, 
e^{-\mathrm{i}\sqrt{2}(\beta_1 x_{\phi_1} + \beta_2 x_{\phi_2})} \, \\ & \times
\chi(\mathrm{i}\beta_1 e^{\mathrm{i}\phi_1}, \mathrm{i}\beta_2 e^{\mathrm{i}\phi_2}).
\end{split}
\end{equation}
Here, $P(x_{\phi_1}, x_{\phi_2})$ denotes the joint distribution of rotated quadratures, which for $\phi_1 = \phi_2 = 0$ reduces to $P(x_1, x_2)$, and for $\phi_1 = 0$, $\phi_2 = \pi/2$ yields the joint position–momentum distribution $P(x_1, p_2)$. Experimentally, the real part $\text{Re}[\chi]$ is extracted by applying two sequential spin-dependent displacements on $\ket{\downarrow}_z\otimes\ket{\Psi}$ and measuring the spin magnetization $\langle\sigma_z\rangle$, while the imaginary part $\text{Im}[\chi]$ is accessed via an additional $\pi/2$ qubit rotation prior to displacement (see Appendix~\ref{app:Experimental_platform}).

In Fig.~\ref{fig2}(b) we show the joint probability distribution $P(x_1, x_2)$  for three different values of the phase $\varphi$. For $\varphi = 0$, the two vdP oscillators are (in-phase) synchronized. For $\varphi = \pi$, they are anti-phase synchronized. In the joint probability distribution, this appears as structures along the diagonal and anti-diagonal directions, respectively. When the phase difference is $\varphi = \pi/2$, we observe a circular shape in the $P(x_1, x_2)$ plot. This can be viewed as the simplest Lissajous figure, formed by two orthogonal oscillators of equal amplitude and a $\pi/2$ phase difference, resulting in a circular trajectory in the plane defined by their position axes (see Appendix~\ref{app:Sync_indicator}). 

We next turn to the quantum regime, where synchronization should persist despite the increased role of quantum fluctuations. To reach this regime, we tune the effective dissipation rates by adjusting the duration of the stroboscopic steps, such that the ratio $\kappa_- / \kappa_+ = 0.42$, leading to steady-state occupation numbers $\bar{n} = 1.4$. The corresponding results are shown in Fig.~\ref{fig2}(c) and (d). Strikingly, even at such low occupation, the system retains the features of limit-cycle behavior: the reconstructed Wigner functions exhibit a visible, albeit smaller, donut-shaped profile; cf. Fig.~\ref{fig2}(c). Moreover, the joint probability distributions continue to display the phase-dependent patterns for $\varphi = 0, \pi/2, \pi$, confirming that synchronization may persist into the quantum regime. Notably, even if only a few quanta per oscillator are present, synchronization between the two modes can be established.

\section{Stability of synchronization to detuning and coupling}

A central feature of classical synchronization is its robustness: even non-identical systems can adjust their rhythms and lock to a common frequency. This behavior manifests in the so-called Arnold tongue~\cite{pikovsky1985universal}, a region in detuning–coupling space where synchronization persists. Analogous regions have been theoretically predicted for coupled quantum limit-cycle systems~\cite{lee2014entanglement, rouletQuantumSynchronizationEntanglement2018a}. However, characterizing this behavior using full state reconstruction is experimentally demanding. To circumvent this challenge, we introduce mutual information extracted from the measurable quadrature distributions $P(x_{\phi_1}, x_{\phi_2})$ as a figure of merit, that retains sensitivity to synchronization~(see Appendix~\ref{app:Sync_indicator}). A measure based on a single quadrature, however, depends on the relative phase difference $\varphi$. For example, at  $\varphi=\pi/2$, $I[x_1:x_2]$ vanishes due to symmetry [middle panel of Fig.~\ref{fig2}(b)], while $I[x_1:p_2]$ establishes that the two oscillators are actually synchronized. Conversely, at $\varphi = 0$, $I[x_1:p_2]$ vanishes and the correlations are contained in $I[x_1:x_2]$. For intermediate phases, both quantities contribute. To eliminate this dependence on $\varphi$, we combine the two, thereby defining a mutual information measure that is invariant under the relative phase:
\begin{equation}
\mathcal{I} = I[x_1:x_2] + I[x_1:p_2],
\end{equation}
where $I[x_1:x_2]$ and $I[x_1:p_2]$ are computed from the respective distributions $P(x_1, x_2)$ and $P(x_1, p_2)$. This serves as our figure of merit for synchronization between the two oscillators.

Similarly to classical synchronization, increasing the dissipative coupling strength $V$ enhances synchronization in the quantum regime. This trend is evident in Figs.~\ref{fig:Arnoldtongue}(a) and (d) which show $\mathcal I$ for zero frequency detuning between the oscillators in the near-classical and quantum regimes, respectively. Note that the  noise of $P(x_1, x_2)$ and $P(x_1, p_2)$, stemming predominantly from quantum projection noise and Rabi frequency fluctuations from the read-out, adds an offset to the mutual information, such that even for zero coupling, $\mathcal I$ has a small finite value.  Conversely, for fixed coupling strength, increasing the detuning $\Delta \omega$ through changing the frequency of $M_2$ progressively suppresses synchronization as the frequency mismatch eventually becomes too large for the interaction to induce phase locking. This behavior is illustrated in Figs.~\ref{fig:Arnoldtongue}(b) and (e) for a fixed effective coupling strength of $V/V_0 = 1.0$. 
To map out synchronization across the full coupling–detuning parameter space, we performed simulations spanning a wide range of parameters [see Figs.~\ref{fig:Arnoldtongue}(c) and (f)]. The simulation results are plotted together with experimental data along the two representative cuts (indicated by the blue dots), including a global offset representing the measured contribution of the read-out noise. 
The Arnold-tongue contour 
shows that synchronization of the two oscillators requires a finite coupling $V$. This arises from the requirement that dissipative interactions must overcome individual damping, which tends to erase the phase information~\cite{lee2014entanglement}.

\section{Phase locking and desynchronization under an external drive \label{sec:ExternalDrive}}

\begin{figure}[ht]
    \centering
    \includegraphics[width=0.9\linewidth]{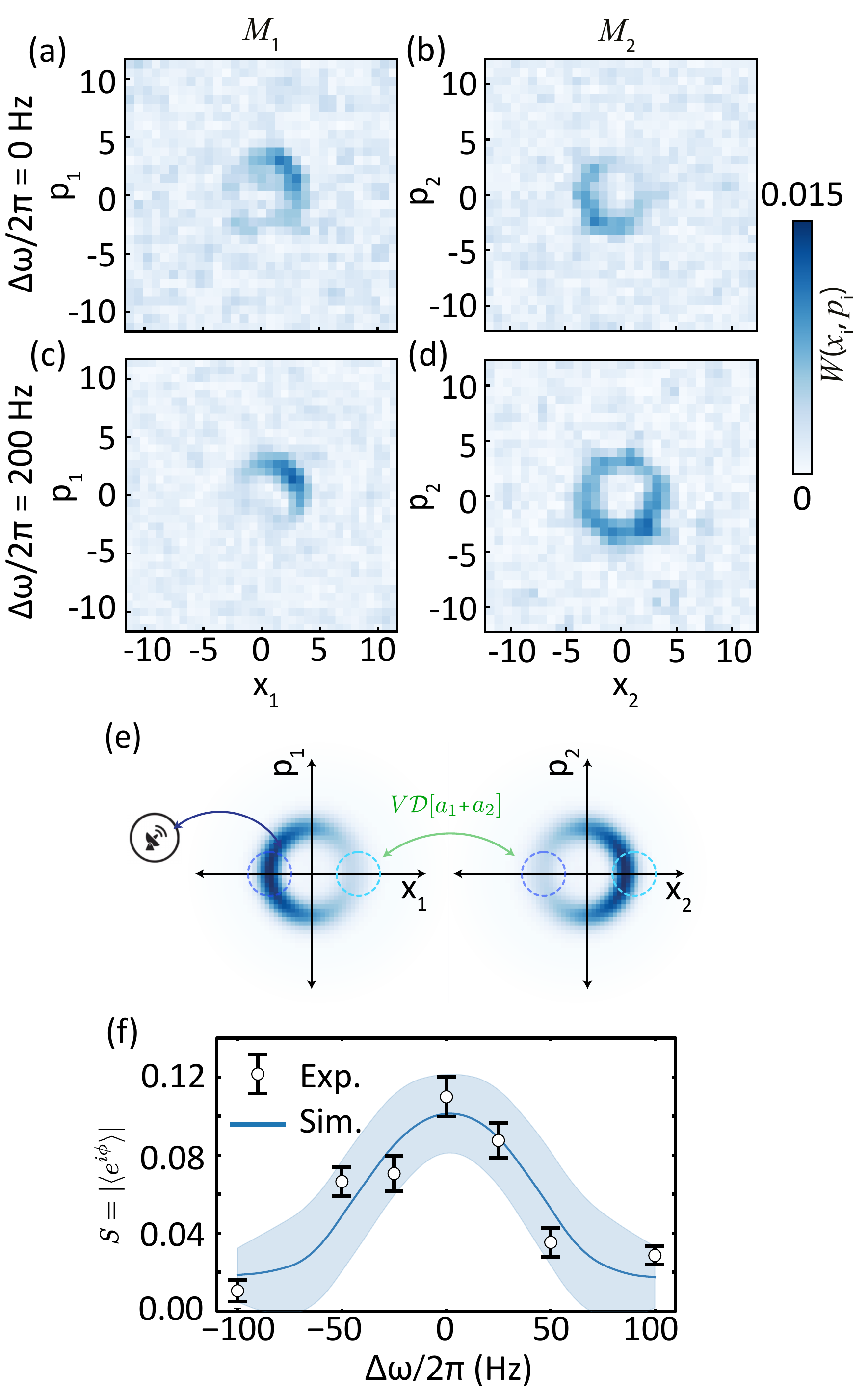}
    \caption{Entrainment of mutual synchronization to an external drive. $M_1$ is driven by an external signal such that its phase aligns with the drive. The second mode $M_2$ is dissipatively locked to $M_1$ with a phase difference of $\varphi = \pi$. (a)(b) Reconstructed Wigner functions of $M_1$ and $M_2$ when the second oscillator is resonant with the external drive. (c)(d)  Same as in (a)(b)  but with the second oscillator detuned by 200\,$\rm Hz$. (e) Illustration of mutual synchronization between $M_1$ and $M_2$ with external drive. (f) Degree of phase localization of the second oscillator, quantified by the mean resultant length $S=|\braket{e^{i\phi}}|$ as a function of frequency detuning $\Delta \omega$ between the oscillator and the external signal. The black circles are experimentally measured $S$, with error bars representing the quantum projection noise. the blue curve is the numerical simulation with the shaded region indicating the uncertainty to motional frequency  and Rabi frequency fluctuations. 
 }
    \label{fig:Sensing}
\end{figure}

We augment the model in Eq.~(\ref{eq:EffectiveModel}) with an additional electric field resonant with $M_1$. This external drive explicitly breaks the underlying $U(1)$-symmetry of the system and imposes a phase preference on the Wigner functions, determined by the drive phase, its amplitude, coupling strengths, and detunings. The resulting configuration can be viewed as a primary–secondary system: the externally driven mode $M_1$ locks to the drive, while the undriven mode $M_2$ synchronizes with $M_1$ through dissipative coupling. By varying the frequency of $M_2$, we can now directly probe the desynchronization process by evaluating the local Wigner functions of each oscillator because of the externally imposed phase preference (as opposed to the previous case without external drive). Figures~\ref{fig:Sensing}(a) and (b) show the Wigner functions of both modes when the frequency of $M_1$ is tuned to match that of the external drive. In this regime,  synchronization emerges, and both modes acquire well-defined phase preferences with the expected phase difference $\varphi = \pi$, set by the phase of the collective dissipation. As the frequency of $M_2$ increases, detuning causes it to desynchronize, while $M_1$ remains phase-locked to the drive [ Figs.~\ref{fig:Sensing}(c) and (d)]. To quantify the degree of phase locking, we use the mean resultant length $S = |\braket{e^{\mathrm{i}\phi}}|$~\cite{mokSynchronizationBoostSinglephoton2020a} of $M_2$. Figure~\ref{fig:Sensing}(f) shows $S$ as a function of the detuning $\Delta\omega$ between $M_1$ and $M_2$, which is maximized at zero detuning and decreases rapidly as the detuning increases and the two modes desynchronize with each other.

Synchronization in this primary–secondary configuration is related to a setup that has been used to enhance the resolution of a classical MEMS-based accelerometer~\cite{MEMSSync}, where a sensing resonator and a readout resonator are coupled unidirectionally. Although in our case the coupling between the two oscillators is reciprocal, we demonstrate that the externally applied drive frequency can nonetheless be estimated by measuring $M_2$. This suggests  opportunities for future work exploring sensing using quantum vdP oscillators ~\cite{vaidyaQuantumSynchronizationDissipative2024a,dutta2019critical} for enhanced resolution or noise suppression analogous to classical synchronized sensors~\cite{PhysRevLett.112.014101}.

\section{Conclusions and Outlook}
Unlike in classical systems, synchronization between quantum systems must contend with the effects of quantum noise and measurement back-action. Here, we have realized the long-anticipated synchronization between two quantum van der Pol oscillators, enabled by engineered collective dissipation in a trapped-ion platform. The synchronized dynamics is encoded in a correlated joint measurement signal and hidden from local measurements, illustrating a uniquely quantum aspect of synchronization. 

The van der Pol oscillator has long served as the archetypal model of classical synchronization, providing the foundation for understanding more complex nonlinear networks. Our results establish its quantum counterpart as a practical and versatile building block, from which richer collective phenomena can be explored. Scaling to larger oscillator networks could uncover genuinely many-body effects beyond mean-field descriptions, such as quantum chimera states~\cite{bastidasQuantumSignaturesChimera2015a} or topologically protected synchronization~\cite{wachtlerTopologicalSynchronizationQuantum2023b}, where classical simulations become intractable and quantum simulators provide a platform to explore these open-system dynamics at scale. Our approach is readily scalable to larger networks of coupled van der Pol oscillators using ion chains. Notably, using a single $\mathrm{^{40}Ca^+}$ ion and $N-1$ $\mathrm{^{44}Ca^+}$ ions~\cite{clos2016time} can unlock the control of $3N$ motional modes, opening the door to programmable oscillator networks. As in classical settings where phase-locked lasers boost output power and coupled sensors enhance sensitivity~\cite{PhysRevLett.112.014101}, analogous advantages may emerge in the quantum regime~\cite{PhysRevE.101.020201, murtadhoCooperationCompetitionSynchronous2023b}.

More broadly, the versatility of trapped ions in combining spin-spin, spin-boson and boson-boson interactions~\cite{zhang2017observation,gorman2018engineering,whitlow2023quantum} with programmable local and collective dissipation~\cite{behrle2023phonon,so2024trapped,PhysRevLett.128.080503} offers a powerful setting for simulating nonequilibrium quantum matter. Our results thus stimulate further exploration of harnessing dissipation as a resource for both understanding and controlling complex quantum many-body dynamics~\cite{verstraete2009quantum,luschen2017signatures,haack2023probing,PhysRevX.14.011026}.

\begin{acknowledgments}
We thank Christoph Bruder, Zherui Chen, Tony Lee, Eric Lutz, Andrew Jayich, Ting Rei Tan, and Juzar Thingna for helpful discussion and reading of the manuscript. We also acknowledge the early contributions of Bingran You, Thomas Lu, and Nils Ciroth to the experimental setup. This work was supported by the U.S. Department of Energy, Office of Science, Office of Basic Energy Sciences under Awards No. DE-SC0023277. Additional support was provided by the  Berkeley NSF-QLCI Challenge Institute for Quantum Computation. C.W.W. was supported by the Deutsche
Forschungsgemeinschaft (DFG, German Research Foundation), Project No. 496502542 (WA 5170/1-1), and received funding from the European Union’s Horizon Europe research and innovation programme under the Marie Skłodowska-Curie Actions (MSCA) grant agreement No. 101149948. Views and opinions expressed are, however, those of the author(s) only and do not necessarily reflect those of the European Union or the European Research Council. Neither the European Union nor the granting authority can be held responsible for them. 

\end{acknowledgments}

\appendix





\section{Experimental platform and protocols \label{app:Experimental_platform}}

\subsection{Mixed-isotope trapped ion quantum simulator}

We implement synchronization of two quantum vdP oscillators using the two axial motional modes $M_1$, $M_2$ in a crystal of a $\mathrm{^{40}Ca^+}$ and a $\mathrm{^{44}Ca^+}$ ion. We use 397\,nm light ($4^2S_{1/2} \leftrightarrow 4^2P_{1/2}$ transition) and 866\,nm light ($4^2P_{1/2} \leftrightarrow 3^2D_{3/2}$ transition) to Doppler cool the $\mathrm{^{40}Ca^+}$ ion while sympathetically cooling the $\mathrm{^{44}Ca^+}$ ion. In addition, we add a sideband 800\,MHz red-detuned from the $4^2S_{1/2} 
\leftrightarrow 4^2P_{1/2}$ transition of the $\mathrm{^{40}Ca^+}$ ion to mitigate collision induced melting of the mixed-isotope crystal~\cite{you2024temporally}. The 866\,nm light passes through an additional electro-optic modulator operated at 4.5\,GHz to provide repumping on $\mathrm{^{44}Ca^+}$ ion to facilitate loading of the $\mathrm{^{44}Ca^+}$ ion and maintaining the ion crystal. We perform single-shot readout of the qubit state on only the $\mathrm{^{40}Ca^+}$ ion with a camera by shining the 397\,nm and 866\,nm light resonantly with the corresponding transition in $\mathrm{^{40}Ca^+}$. From the camera image, the position of the $\mathrm{^{40}Ca^+}$ in the ion crystal can be extracted to detect switching of the $\mathrm{^{40}Ca^+}$-$\mathrm{^{44}Ca^+}$ crystal orientation due to collisions with the background gas. If such an event is detected, we lower the trap rf power and raise it back again until the desired order is restored. 

$M_1$, $M_2$ are the in-phase and out-of-phase axial modes, whose frequencies are 522\,kHz and 906\,kHz, respectively. After Doppler cooling, we perform sideband cooling on both modes to prepare them close to the ground state. The qubit states we use are  $\ket{\downarrow}_z = 4^2S_{1/2} \,(\rm m=-1/2)$ and $ \ket{\uparrow}_z =3^2D_{5/2}\,(\rm m=-5/2)$. The qubit is initiated in the $\ket{\downarrow}_z$ state by optical pumping with  $\sigma_-$ polarized 397\,nm light and linear 866\,nm repumping light.  The motional gain, loss and the dissipative coupling between the two oscillators are generated by driving the motional sidebands of the qubit transition of the  $\mathrm{^{40}Ca^+}$ ion $\ket{\downarrow}_z\leftrightarrow \ket{\uparrow}_z$ using a narrow-linewidth 729\,nm laser locked to a ultra-low expansion cavity, followed by qubit reset using 854\,nm light resonant with  $\ket{\uparrow}_z \leftrightarrow 4^2P_{3/2}\,(\rm m=-3/2)$. The latter state mostly decays back to $\ket{\downarrow}_z$. The leakage out of the qubit manifold (from  $4^2P_{3/2}$ to $3^2D_{3/2}$ process) is only 0.6\%~\cite{roos2000controlling} and can be closed with short optical pumping pulse once every five evolution cycles.\\

\subsection{Dissipative and coherent quantum controls}
The 729\,nm light passes through a double-pass acousto-optic modulator (AOM), which is used to change the frequency of the light to match the qubit transition frequency with a constant offset of +80\,MHz. It then passes through a single-pass AOM centered at -80\,MHz before coupling to the ions, which enables control over dissipative and coherent operations on the sidebands. The single-pass AOM is driven by an arbitrary waveform generator (Spectrum M4i.6621-x8). Throughout the experiment, the phase of each RF-signal is referenced to a common clock to ensure phase continuity.  For implementing negative damping (via BSB) and nonlinear damping (via 2RSB), we tune the single-pass AOM to match the frequencies of the corresponding sidebands and randomize the phase of the qubit-motion coupling . In the dissipative coupling, the two RSBs are driven with a fixed phase difference $\varphi$, which sets the phase reference for the mutual synchronization. To read out the motional state, we use the state-dependent force \cite{PhysRevLett.94.153602} whose phase defines the readout axis in phase space. For the phase locking to the external drive, an oscillating electric field at the same frequency of $M_1$ is applied to one of the trap electrodes to excite the ion motion. This drive is controlled by a separate channel of the same arbitrary waveform generator to ensure phase coherence with the optical modulation.\\  

\begin{table*}[ht!]
    \centering
    \begin{tabular}{|c|c|c|c|c|}
    \hline 
 & \multicolumn{2}{|c|}{Near-Classical }& \multicolumn{2}{|c|}{Quantum } \\ \hline 
         &  $\quad \quad M_1 \quad \quad $&  $\quad \quad M_2\quad \quad $&  $\quad \quad M_1\quad \quad $&  $\quad \quad M_2\quad \quad $\\ \hline
 $\eta_i$& 0.094& 0.072& 0.094&0.072\\\hline 
         $\Omega_{\rm BSB}/2\pi\,({\rm MHz})$&  0.12&  0.12&  0.11&  0.11\\ \hline 
         $\Omega_{\rm 2RSB}/2\pi\,({\rm MHz})$&  0.22&  0.22&  0.22&  0.22\\ \hline
 $\tau_{\rm BSB}(\mu s) $& 17.10& 22.46& 17.10& 22.45\\\hline        
 $\tau_{\rm  2RSB}(\mu s)$& 48.77& 84.06& 97.54& 168.12\\\hline
 
 $\kappa_+/2\pi\,({\rm kHz})$& 0.12& 0.12& 0.10& 0.10\\\hline
 $\kappa_-/2\pi\,(\rm kHz)$ & 0.017& 0.017& 0.042& 0.042\\ \hline 

 $\Omega_{\rm sync}/2\pi\,({\rm MHz})$& 0.048& 0.063& 0.048& 0.063\\
 \hline
  $\tau_{\rm  sync}(\rm \mu s)$& \multicolumn{2}{|c|}{24}& \multicolumn{2}{|c|}{32}\\ \hline 
 $V_0/2\pi\,(\rm kHz)$& \multicolumn{2}{|c|}{0.1}& \multicolumn{2}{|c|}{0.1}\\ \hline
    \end{tabular}
    \caption{The experimental parameters for synchronization of two vdP oscillators in the near-classical and quantum regimes. }
    \label{Table_VdP_parameters}
\end{table*}

\subsection{Readout of the motional states} 
\noindent Here we discuss the reconstruction of the Wigner function of a single motional mode $W(x,p)$ and the reconstruction of the joint probability function $P(x_1,x_2)$ [$P(x_1,p_2)$] between two motional modes [Reconstruction measurement in Fig.~\ref{fig_ex:quantum_circuit}(a)]. To reconstruct the Wigner function of a single target motional state $\ket{\psi_{osc}}$, we read out the characteristic function by applying state-dependent forces (SDF) with different time duration and phases, which we couple to the RSB and BSB of the same mode simultaneously with phases $\phi_r$ and $\phi_b$ . In the readout process, we apply the SDF in the $\sigma_x$ basis by keeping $\phi_b+\phi_r = \pi$ with Hamiltonian
\begin{equation}
    H_{\mathrm{SDF}} = \frac{\eta\Omega}{2}\sigma_x(ae^{\mathrm{i}\phi_m}+ a^\dagger e^{-\mathrm{i}\phi_m}).
\end{equation}
Here, $\eta$ is the Lamb-Dicke parameter and $\Omega$ is the Rabi frequency of the SDF. The motional phase $\phi_m = (\phi_r-\phi_b)/2$ is controlled by the relative phase of the blue and red sideband. The initial qubit-motion state before applying the SDF is $\ket{\psi(0)} = \ket{\psi_{\rm osc}} \otimes \ket{\downarrow}_z$. During time evolution with $H_{\rm{SDF}}$, the motional state is coupled to the $\sigma_x$ basis states, $\ket{\uparrow}_x$ and $\ket{\downarrow}_x$, displacing the oscillator towards opposite directions in phase space. The state after applying SDF for time $t$ is
\begin{equation}
    \ket{\psi(t)} = \frac{1}{\sqrt{2}}(\ket{\uparrow}_x\mathcal{D}(\beta)\ket{\psi_{\rm osc}}+\ket{\downarrow}_x\mathcal{D}(-\beta)\ket{\psi_{\rm osc}}).
\end{equation}
Here, $\mathcal{D}(\beta)=\exp(\beta a^\dagger-\beta^*a)$ is the displacement operator with $\beta = \frac{\mathrm{i}}{2}\eta\Omega te^{\mathrm{i}\phi_m}$. By measuring in the $\sigma_z$ basis we obtain the real part of the characteristic function
\begin{equation}
    \braket{\sigma_z} = \mathrm{Re} [\bra{\psi_{\rm osc}}\mathcal{D}(2\beta)\ket{\psi_{\rm osc}})] = \mathrm{Re} [ \chi(2 \beta)].
\end{equation}
To read out the imaginary part, we prepare the initial state for in the $\sigma_y$ basis $\ket{\psi(0)} = \ket{\psi_{\rm osc}} \otimes \ket{\downarrow}_y$. The subsequent SDF drive and measurement in the qubit $\sigma_z$ basis are the same as the readout of the real part. After obtaining the real and imaginary part of the characteristic function, we then apply the two dimensional Fourier transform and reconstruct the Wigner function
\begin{equation}
W(\alpha) = \frac{1}{\pi^2} \int d^2\beta \; \chi(2\beta) \; e^{2\mathrm{i}(\beta_r \alpha_i - \beta_i \alpha_r)}.
\end{equation}

In addition, quantum projection noise, state preparation, and other imperfections introduce a global offset of the qubit population. We estimate and remove the constant offset of the characteristic function by averaging over its marginal distribution~\cite{fluhmann2020direct}. After subtracting this offset, we apply zero-padding to enhance the spatial resolution of the reconstructed Wigner function.

For reconstructing the joint probability distribution $P(x_1,x_2)$ ($P(x_1,p_2)$ or other combinations), we apply two subsequent SDFs on the two motional modes as shown in Eq.~(\ref{eq_readout}). Similar to the single mode Wigner function, reading out the real part of the joint characteristic function requires starting in $\ket{\downarrow}_z$ while reading out the imaginary part requires starting in $\ket{\downarrow}_y$. By changing the motional phases on the SDFs on the two modes, we are able to control the readout axes of the two mode joint probability distribution. 

.
\begin{figure*}
    \centering
    \includegraphics[width=1\linewidth]{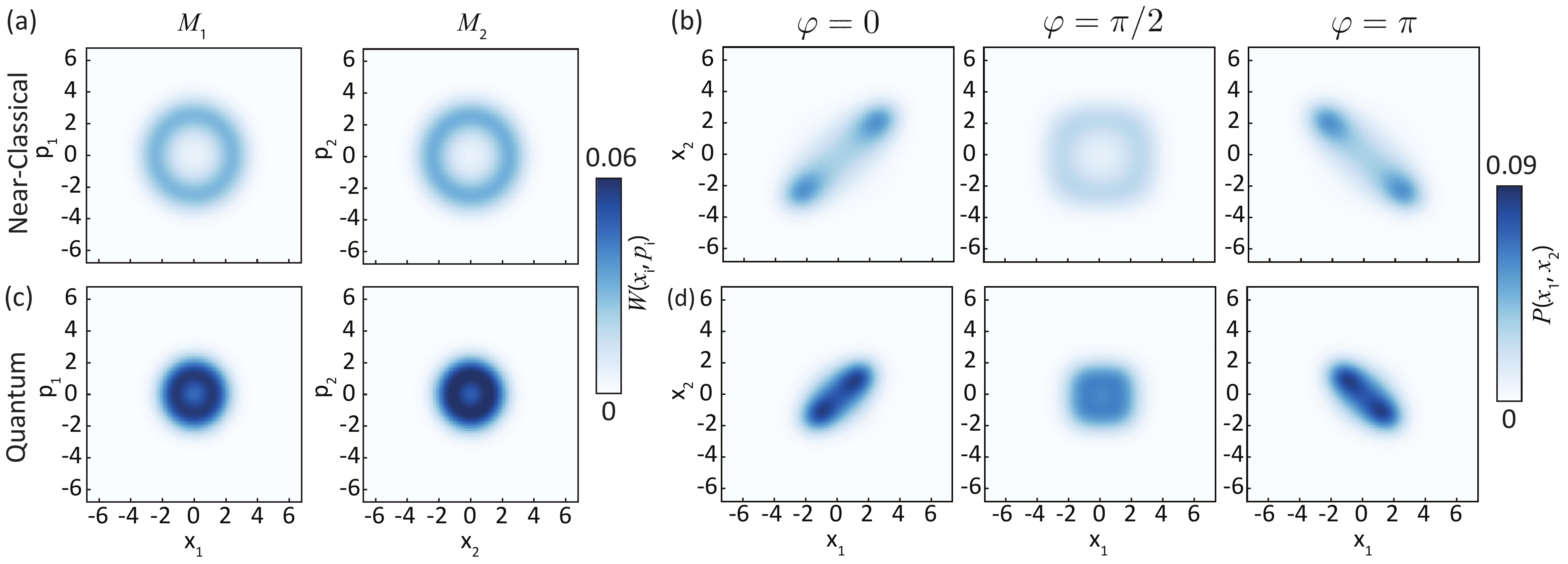}
    \caption{ Numerical simulation of synchronization between two vdP oscillators in the near-classical and quantum regimes.   (a)(b) Near-classical regime with $\kappa_- / \kappa_+ = 0.14$ and mean phonon number $\bar{n} = 3.5$. 
(a) Simulated Wigner functions show the expected donut-shaped profiles without phase preference.
(b) Joint probability distributions reproduce in-phase, circular and anti-phase correlations for  $\varphi = 0$,  $\pi/2$, and $\pi$, respectively. 
(c)(d) Same as panels (a)(b) but in the quantum regime with $\kappa_- / \kappa_+ = 0.42$ and $\bar{n} = 1.4$. 
These simulations correspond to the experimental results of Fig.~2.
    }
    \label{fig_SI_Px1x2}
\end{figure*}

\section{Experimental parameters and calibration procedures \label{app:Experimental_parameters}}

\subsection{Experimental parameters for synchronizing two vdP oscillators.}

The synchronization experiment is conducted mainly in two parameter regimes, with vdP oscillators  operating in the near-classical ($\overline{n} = 3.5$) and quantum ($\overline{n} = 1.4$) regime \cite{mokSynchronizationBoostSinglephoton2020a}. To generate the two vdP oscillators with similar occupation number, we tune the nonlinear and negative damping strengths such that $\kappa_-/\kappa_+$ are identical for both oscillators. Experimentally, for each of the sideband operations, we keep the same laser power and adjust the sideband drive time to account for the different Lamb-Dicke parameters of the two modes to achieve similar damping strength. For example, the duration of the blue sideband (BSB, $\ket{\downarrow}\ket{n}\leftrightarrow\ket{\uparrow}\ket{n+1}$ transition) drive is $\tau_{\rm BSB}\propto 1/\eta_i$, while the duration of the second order red sideband (2RSB, $\ket{\downarrow}\ket{n}\leftrightarrow\ket{\uparrow}\ket{n-2}$ transition) drive is $\tau_{\rm 2RSB}\propto 1/\eta^2_i$, where $\eta_i$ is the Lamb-Dicke parameter for mode $M_i$ of the $\mathrm{^{40}Ca^+}$. In Table~\ref{Table_VdP_parameters} we list the experimental parameters used for the synchronization experiment in the two regimes: Lamb-Dicke parameters $\eta_i$ for the two modes; corresponding carrier Rabi frequencies of the sideband operations $\Omega_{\rm{BSB}}$ and $\Omega_{\rm{2RSB}}$ to generate effective dampings, the duration $\tau_{\rm{BSB}},\,\tau_{\rm{2RSB}}$ of applying the BSB and 2RSB drives and the resulting effective rates $\kappa_+,\,\kappa_-$; the corresponding carrier Rabi frequencies of simultaneous RSB drives $\Omega_{\rm sync}$ on the two modes to generate the dissipative coupling with interaction time $\tau_{\rm sync}$ and resulting effective coupling rate $V_0$. 

In Fig.~\ref{fig_SI_Px1x2}, we show the numerical simulation using QuTiP~\cite{JOHANSSON20121760}\ for synchronization between two vdP oscillators in near-classical and quantum regimes using the parameters from Table~\ref{Table_VdP_parameters}, which shows good agreement with the experimental results. \\

\begin{figure*}
    \centering
    \includegraphics[width=0.8\linewidth]{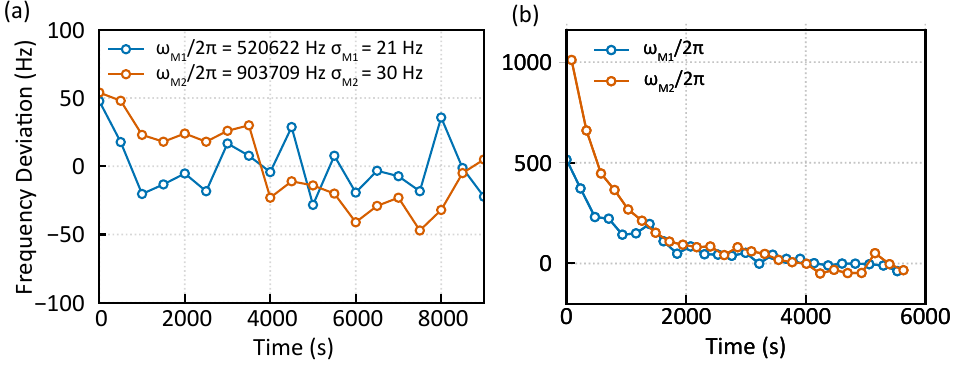}
    \caption{(a) The measured relative frequencies of axial modes $M_1$ (blue) and $M_2$ (orange) during a typical data collection run.  (b) The frequency relaxation process of the two modes upon switching from low rf power (for loading ions) to high rf power (for experiments) due to thermal expansion of the trap rf electrodes.}
    \label{fig_motion_frequency}
\end{figure*}

\subsection{Frequency calibration of qubits and the motional modes.} 
During data collection, we recalibrate the qubit frequency every 180\,s to compensate for slow drifts in the reference cavity and magnetic field, and we recalibrate the motional frequencies of the two axial modes every 400\,s.

Long-term stability of the motional frequencies is essential, as
both the synchronization strength and the motional reconstructions are sensitive to
frequency drift. For the motional frequency calibration, we simultaneously probe the RSBs and BSBs on both modes using low laser intensity and an interrogation time of approximately 2\,ms to reduce linewidth and laser induced ac-Stark shifts~\cite{PhysRevLett.90.143602}. To reduce the impact of qubit frequency fluctuations during the motional frequency calibration, we interleave the measurements by collecting one data point from BSB and one data point from RSB in a repeating cycle. The motional frequencies are then extracted from the difference between the BSB and RSB fit and updated for the subsequent experiments. Figure~\ref{fig_motion_frequency}(a) shows the relative frequency drift of modes $M_1$ and $M_2$ over a two-hour interval after the trap has reached thermal equilibrium with its environment; the residual fluctuations are small, with standard deviations $\sigma_{M_1}=21\,\mathrm{Hz}$ and $\sigma_{M_2}=30\,\mathrm{Hz}$. Remaining noise likely originates from the sources providing the endcap electrodes voltages, environmental electric-field noise, and uncertainties in the sideband frequency measurements. These frequency uncertainties are included in the numerical simulations of mutual synchronization and phase locking to the external drive.

By contrast, the motional mode frequencies drift much more rapidly immediately after loading the two-ion crystal [Fig.~\ref{fig_motion_frequency}(b)]. During loading, the trap is operated at reduced RF power to facilitate trapping of the mixed-isotope chain; after switching to the higher RF power used for the synchronization measurements, resistive heating of the RF electrodes leads to thermal expansion, which increases the effective endcap separation and reduces the trap frequency. The frequency then relaxes toward a steady value on a timescale of one hour.\\

\subsection{Phase control in mutual synchronization.}

\begin{figure}[]
    \centering
    \includegraphics[width=\linewidth]{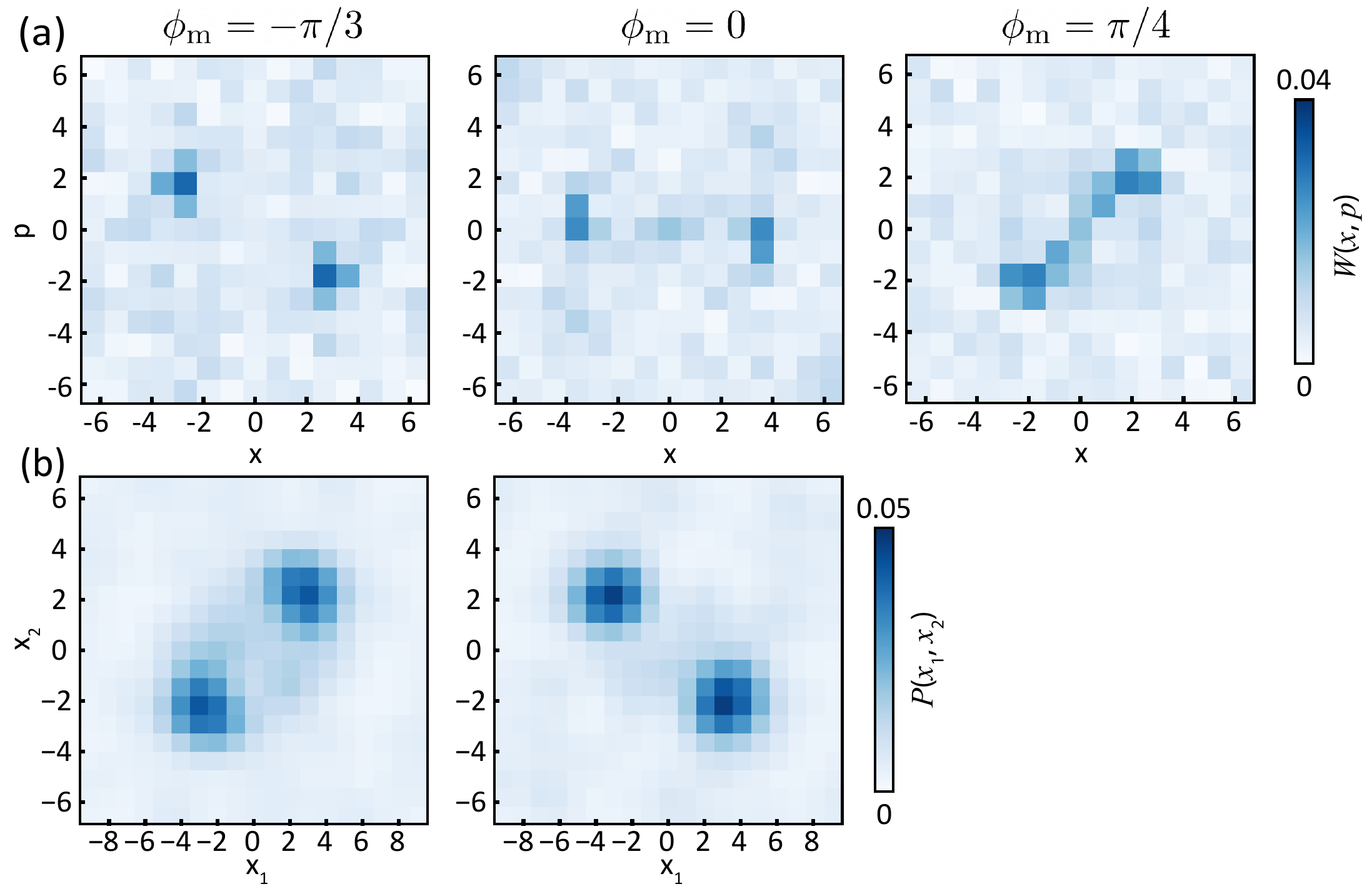}
    \caption{(a) Wigner functions of $\varrho=\frac{1}{2}(\ket{\alpha}\bra{\alpha}+\ket{-\alpha}\bra{-\alpha})$ with motional phase $\phi_{\rm m}=-\pi/3, 0$ and $\pi/4$ with $60$\,$\rm {\mu s}$ SDF drive and $30$\,$\rm {\mu s}$ readout time. The corresponding carrier Rabi frequency of the SDF drive is $\Omega/2\pi= 0.077$\,MHz. (b) Joint probability distribution $P(x_1,x_2)$ for the mixture of opposite two-mode coherent-state  on $M_1$ and $M_2$ ($\varrho=\frac{1}{2}[\varrho(\alpha_1,\alpha_2)+\varrho(-\alpha_1,-\alpha_2)]$) with a phase difference of $0$ and $\pi$. 
    }
    \label{fig_SI_motion_phase}
\end{figure}

With the precise frequency calibration of the motional modes, we then demonstrate the capability to accurately control the motional phases . We prepare the state in $\ket{\psi}=\ket{0}\otimes\ket{\downarrow}_z$ and drive the motional mode with the state dependent force (SDF) $H_{\mathrm{SDF}} = \frac{\eta\Omega}{2}\sigma_x(ae^{\mathrm{i}\phi_m}+ a^\dagger e^{-\mathrm{i}\phi_m})$. The SDF entangles the qubit and motional degree of freedom, and the system evolves into a cat state~\cite{PhysRevLett.94.153602}. After resetting the qubit using repump light, the system is in a mixed state of two coherent states $\varrho_{\rm osc}=\frac{1}{2}\ket{\alpha}\bra{\alpha}+\frac{1}{2}\ket{-\alpha}\bra{-\alpha}$, where  $\alpha=\frac{1}{2}\eta\Omega te^{\mathrm{i}\phi_m}$ is the displacement after applying the SDF for time $t$. Then the motional mode is measured with the protocol described in the previous section on experimental protocols. In Fig.~\ref{fig_SI_motion_phase}(a), we show the  Wigner functions corresponding to the motional phases $\phi_m=-\pi/3, 0, \pi/4$,  which align well with the expected results. We also illustrate the coherent phase control between the two motional modes $M_1$ and $M_2$. Specifically, we apply the SDF to create the two-mode cat state $\ket{\psi}=\frac{1}{\sqrt{2}}(\ket{\uparrow}_x \ket{\alpha_1,\alpha_2} + \ket{\downarrow}_x \ket{-\alpha_1,-\alpha_2})$, which after qubit-reset is projected onto the mixed state  $\varrho_{\rm osc}=\frac{1}{2}[\varrho(\alpha_1,\alpha_2)+\varrho(-\alpha_1,-\alpha_2)]$. When the displacement $\alpha_1$ of $M_1$ is along the position axis $(\phi_{\rm m1} = \pi/2)$ and the displacement of $M_2$ is also aligned with the position axis $(\phi_{\rm m2} = 0, \pi)$, the joint probability distribution $P(x_1, x_2)$ exhibits two lobes oriented along the diagonal and antidiagonal directions [Fig.~\ref{fig_SI_motion_phase}(b)].\\

\subsection{Calibration of external drive}
\begin{figure}
    \centering
    \includegraphics[width=1\linewidth]{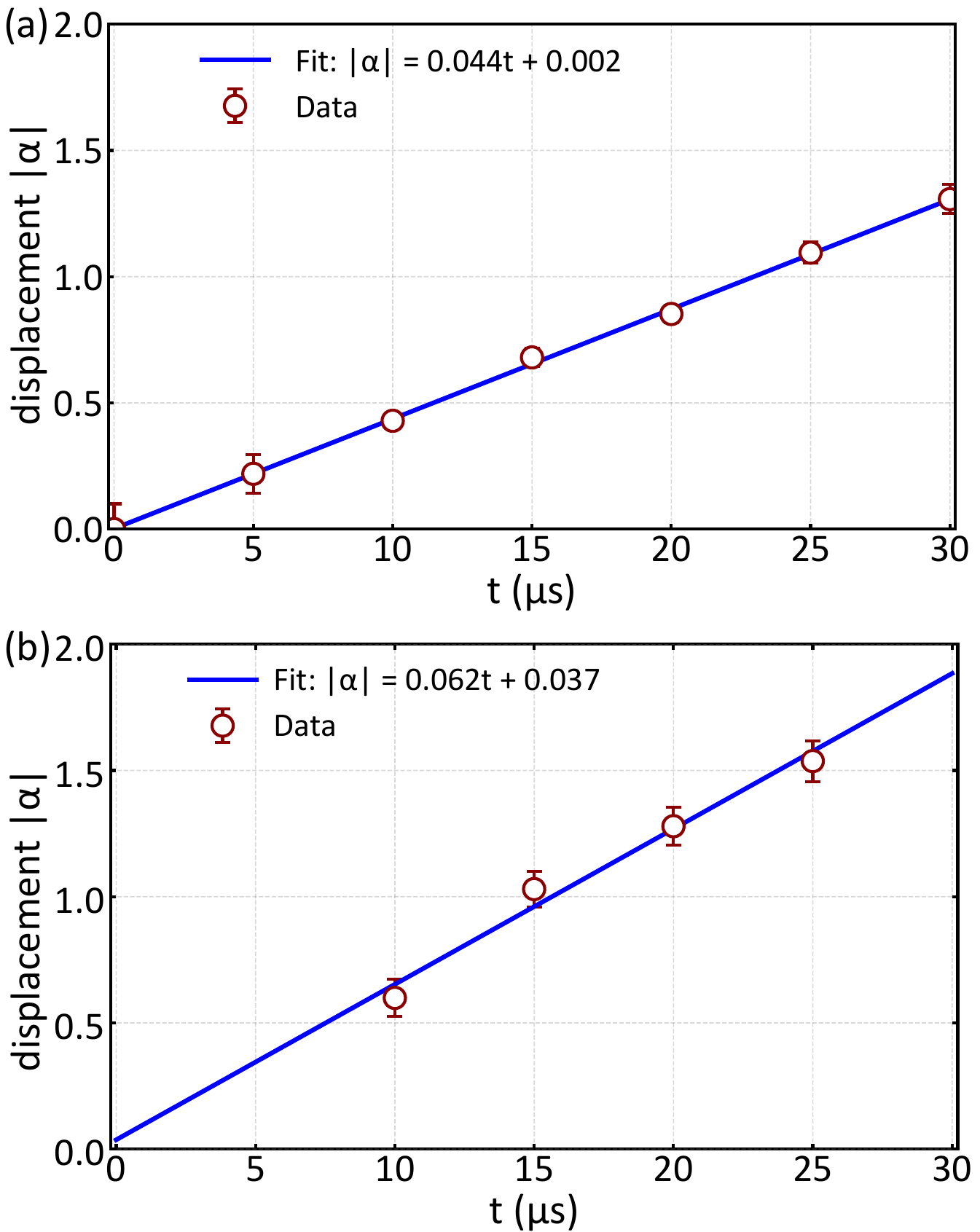}
    \caption{Fitted displacement $|\alpha|$ from BSB flopping as a function of the time $t$ we apply the external field with constant amplitude. (a) Single $^{40}{\rm Ca}^+$ ion axial mode. (b) $^{40}{\rm Ca}^+$-$^{40}{\rm Ca}^+$ in-phase mode ($M_1$).} 
    \label{fig_external_drive_calibration}
\end{figure}

By driving the ion crystal with an oscillating electrical field applied on one of the endcap electrodes, the motional mode will be coherently excited~\cite{PhysRevLett.76.1796} and the corresponding Hamiltonian is
\begin{equation}
    H=\Omega_{d}(a e^{\mathrm{i}\varphi_{\rm d}}+a^\dagger e^{-\mathrm{i}\varphi_{\rm d}}).
\end{equation}
Here, $\Omega_d$ is the strength of external drive and $\varphi_d$ is the phase of the external drive. With this additional external drive, the phase of the vdP oscillator will be entrained to the external drive. The  resulting Wigner function breaks the radial symmetry and is attracted towards the drive axis (see Fig.~\ref{fig_ex:external_drive}). To calibrate the strength of the external drive, we apply the oscillating electric field resonant with a specific motional mode for different amount of time and use the blue sideband flopping to probe the motional occupation. After the system evolved for time $t$, the displacement induced by the external drive is $\alpha=\Omega_d te^{\mathrm{i}\varphi_d}$. To extract the amplitude of displacement $|\alpha|$, we fit the blue sideband Rabi flopping assuming a coherent displacement of a thermal state~\cite{RevModPhys.75.281}. For the same external drive amplitude, the coupling strength for the axial mode of a single ion is found to be $\Omega_{d}/2\pi = 6.94$\,kHz, while the coupling strength to the in-phase mode ($M_1$) of the $^{40}{\rm Ca}^+$–$^{44}{\rm Ca}^+$ ion crystal is found to be $\Omega_{d,M_1}/2\pi = 9.80$\,kHz (Fig.~\ref{fig_external_drive_calibration}). Their ratio of 1.41 is consistent with the theoretical calculation $\Omega_{d,M_1}/\Omega_{d}=1.43$. Note that for a pure $^{40}{\rm Ca}^+$-$^{40}{\rm Ca}^+$ ion crystal, the ratio is $\sqrt2$.

\section{Error analysis in stability of synchronization}
\label{app:Error_analysis}
To further analyze the stability of synchronization to detuning and coupling strength, we perform numerical simulations that incorporate motional frequency drifts and Rabi frequency fluctuations from intensity noise and beam pointing errors. Specifically, we compute the mutual information as a function of frequency difference $\Delta\omega$ and coupling strength $V$: $\mathcal{I}(\Delta\omega, V)$. The coupling strength fluctuation $\delta V$ and frequency fluctuation $\delta \omega$ contribute to uncertainty of mutual information. We calculate this uncertainty from the standard deviation of 20 numerical simulations using Gaussian distributed coupling strengths with standard deviation 2\% and frequencies with standard deviation $\Delta \omega/2\pi=\sqrt{\sigma_1^2+\sigma_2^2} =37\,\text{Hz}$. The estimation of the mutual information uncertainty interval $[\mathcal{I}-\Delta \mathcal{I},\mathcal{I}+\Delta \mathcal{I}]$ is plotted as the shaded area in the main text Fig.\ref{fig:Arnoldtongue}(a)-(e). The error bar on the data reflects quantum projection noise on $\mathcal{I}$, which is calculated by adding random quantum projection noise to the simulation and computing the standard deviation of 20 samples of noisy mutual information. We use a similar method for analyzing experimental imperfections and statistical errors of $S= |\braket{e^{i\phi}}|$ in Fig.~\ref{fig:Sensing}(f).

\section{Effective interaction generated with sideband transition \label{app:Effective_interaction}}

\subsection{Derivation of effective dissipation.} 
To obtain the effective dynamics on the oscillators (Eq.\,1), we begin by deriving the effective Lindblad dissipator on the bosonic modes from a qubit-motion coupling Hamiltonian. The most general form of the qubit-motion coupling is 
\begin{equation}\label{eq_spin_motion}
    H=\Omega(\sigma_+{O}+\sigma_-{O}^\dagger),
\end{equation}
where ${O}$ represents a certain bosonic operator (e.g. $a^\dagger$ for the BSB, $a^2$ for the 2RSB and $a$ for the RSB), $\Omega$ is the generalized Rabi frequency of the sideband transition, and $\sigma_+$ ($\sigma_-$) is the qubit raising (lowering) operator. At the beginning of each step, the qubit state is reset to $\ket{\downarrow}_z$, so that the qubit-motion density matrix takes the form
\begin{equation}
    \varrho =\ket{\downarrow}_z\bra{\downarrow}_z \otimes\varrho_{\rm osc}.
\end{equation}

\begin{figure*}[hbt!]
    \centering
    \includegraphics[width=0.95\linewidth]{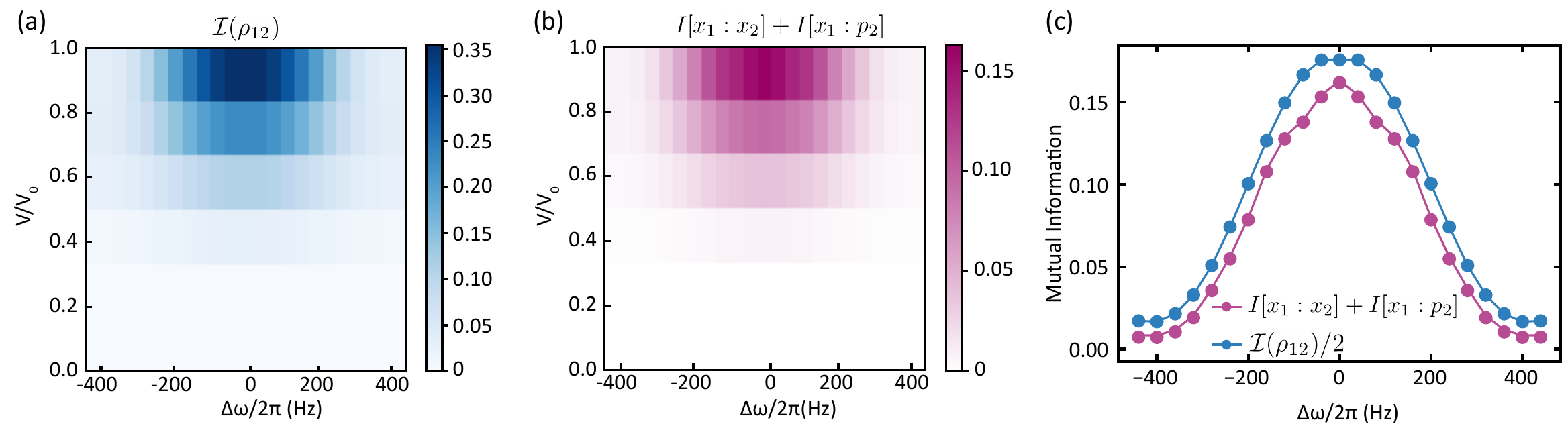}
    \caption{(a) Mutual information between two motional modes as a function of coupling strength and frequency detuning. (b) Summation of mutual information between the quadratures $x_1$--$x_2$ and $x_1$--$p_2$. (c) Comparision between $\mathcal{I}(\varrho_{12})/2$ and $I[x_1:x_2]+I[x_1:p_2]$ as a function of frequency detuning. }
    \label{fig_ex:MutualInfo}
\end{figure*}

After a coherent evolution under $H$ for a small duration $\delta t$, 
followed by another qubit reset, the motional state is obtained by taking the partial trace of $\varrho$ over the qubit degree of freedom
\begin{equation}
    \varrho_{\rm osc}(\delta t)= \text{Tr}_\mathrm{s}\left[ e^{-\mathrm{i} \hat{H} \delta t} \varrho e^{\mathrm{i} \hat{H} \delta t}\right]=\sum_{s=\downarrow, \uparrow}\braket{s|e^{-\mathrm{i}\hat{H}\delta t}\varrho_{\rm osc} e^{\mathrm{i}\hat{H}\delta t}|s}.
\end{equation}
Expanding each term up to second order in $\delta t$ and neglecting higher-order corrections, the $\ket{\downarrow}_z$ and $\ket{\uparrow}_z$ components of the trace are given by $\braket{\downarrow|\varrho(\delta t)|\downarrow}=\varrho_{\rm osc}-\frac{1}{2}\Omega^2\delta t^2\{O^\dagger O, \varrho_{\rm osc}\}$ and $\braket{\uparrow|\varrho(\delta t)|\uparrow}=\Omega^2\delta t^2\hat{O}\varrho_{\rm osc}\hat{O}^\dagger$. The density matrix of the motional degrees of freedom after a time $\delta t$ is given by 
\begin{equation}
    \varrho_{\rm osc}(\delta t)\approx\varrho_{\rm osc}+\Omega^2\delta t^2\hat{O}\varrho_{\rm osc}\hat{O}^\dagger-\frac{1}{2}\Omega^2\delta t^2\{O^\dagger O, \varrho_{\rm osc}\}.
\end{equation}
In the limit $\delta t \to 0$ the time derivative of the reduced density matrix is then given by 
\begin{equation}
\dot \varrho_\text{osc} (t)= \lim\limits_{\delta t \to 0} \frac{\varrho_\text{osc}(t+\delta t) - \varrho_\mathrm{osc}(t)}{\delta t} = \kappa ~ \mathcal D[\hat{O}] \varrho_\text{osc} (t),
\end{equation}
where $\mathcal{D}[\hat{O}]$ denotes a collapse operator acting on the density matrix, i.e.,  $\mathcal{D}[\hat{O}]\varrho=\hat{O}\varrho\hat{O}^\dagger-\frac{1}{2}\{\hat{O}^\dagger \hat{O},\varrho\}$. The resulting effective damping rate is then given by  $\kappa=\Omega^2\delta t $ . We now recall that  the Hamiltonian of the BSB and 2RSB transitions are~\cite{RevModPhys.75.281}:
\begin{equation}
\begin{aligned}
    H_{\rm BSB}^{(i)} &= \frac{\mathrm{i}}{2}\,\Omega_{\rm BSB}^{(i)}\eta_i \left( \sigma_+ a_i^\dagger e^{\mathrm{i}\varphi_b} - \sigma_- a_i e^{-\mathrm{i}\varphi_b} \right), \\
    H_{\rm 2RSB}^{(i)} &= -\frac{1}{4}\,\Omega_{\rm 2RSB}^{(i)}\,\eta_i^2 \left( \sigma_+ a_i^2 e^{\mathrm{i}\varphi_r} + \sigma_- a_i^{\dagger 2} e^{-\mathrm{i}\varphi_r} \right).
\end{aligned}
\end{equation}
Here, $H_{\rm BSB}^{(i)}$ and $H_{\rm 2RSB}^{(i)}$ denote the interaction Hamiltonians for the 
BSB and 2RSB transitions on mode $M_i$, respectively. $\Omega_{\rm BSB}^{(i)}$ and $\Omega_{\rm 2RSB}^{(i)}$ are the corresponding 
Rabi frequencies, $\eta_i$ is the Lamb-Dicke parameter, and $\varphi_b$\,($\varphi_r$) 
denotes the phase of the BSB (2RSB) drive. The annihilation (creation) operators of motional mode $M_i$ are denotes as $a_i$\,($a_i^\dagger$).

Additionally, we define each cycle as the sum of all sideband interaction times 
\begin{equation}  
 T = \sum_{i=1,2} \left(t_{\rm BSB}^{(i)}+ t_{\rm 2RSB}^{(i)}\right)+t_{\rm sync} ,
\end{equation}
where the BSB and the 2RSB interaction times on $M_i$ are $t_{\rm BSB}^{(i)}$ and $t_{\rm 2RSB}^{(i)}$. Consequently, the negative damping $\mathcal{D}[a^\dagger]$ and nonlinear damping $\mathcal{D}[a^2]$ have effective rates  $\kappa_+=\frac{\Omega_{\rm BSB}^2\eta^2t_{\rm BSB}^2}{4T}$ and $\kappa_-=\frac{\Omega_{\rm 2RSB}^2\eta^4t_{\rm 2RSB}^2}{16T}$, respectively.

To realize the collective dissipation that enables the synchronization, we simultaneously drive the RSBs on both oscillators with a relative phase $\varphi$ and equal strength, and then perform a qubit reset. The Hamiltonian of the collective RSBs drive is
\begin{equation}
H_\mathrm{sync} = \mathrm i \frac{\Omega_\mathrm{sync} \eta}{2}\left[\sigma^- \left(a_1^\dagger + e^{-\mathrm i \varphi} a_2^\dagger \right) - \sigma^+ \left(a_1 + e^{\mathrm i \varphi} a_2 \right)\right],
\label{eq:Hsync}
\end{equation}
corresponding to the dissipator $V \mathcal D\left[a_1 - e^{\mathrm i \varphi} a_2 \right]$. Here, for simplicity, we write the equal RSBs strength as $\Omega_\mathrm{sync} \eta = \Omega_\mathrm{sync}^{(1)} \eta_1= \Omega_\mathrm{sync}^{(2)} \eta_2$. Eq.~\ref{eq:Hsync} can be written in the form of general qubit-motion Hamiltonian (Eq.~\ref{eq_spin_motion}) with $\hat{O}=\mathrm{i}(a_1-e^{\mathrm{i}\varphi}a_2)$ and $\Omega=\Omega_{\rm sync}\eta/2$. Following the above derivation, the effective dissipation rate is $V=\frac{\Omega_{\rm sync}^2\eta^2\tau_{\rm sync}^2}{4T}$.\\

\subsection{Frequency tuning between the two oscillators.}
To study the robustness of the synchronization against frequency mismatch between the two vdP oscillators, we tune the frequency difference $\Delta\omega = \delta_2 - \delta_1$ between the two oscillators by changing the frequency of the laser fields. Specifically, to set the frequency of mode $M_i$ to $\delta_i$, we tune the frequencies of the BSB, RSB, and 2RSB drives to $\omega_i +\delta_i$, $-(\omega_i +\delta_i)$ and $-2(\omega_i +\delta_i)$, respectively. 

\section{Synchronization Indicators and Dynamics \label{app:Sync_indicator}}

\subsection{Mutual Information as Indicator of Synchronization.}
Mutual information quantifies the total amount of correlations between two subsystems. Since synchronization, as discussed here, also relies on correlations between the two vdP oscillators, one can expect mutual information to serve as a good measure of (quantum) synchronization~\cite{ameriMutualInformationOrder2015a}. Given the steady-state density matrix of the two oscillators $\varrho_{12}$, the mutual information is defined as
\begin{equation}
\mathcal I(\varrho_{12}) = S(\varrho_{1}) + S(\varrho_{2}) - S(\varrho_{12}),
\end{equation}
where $S(\varrho) = -\mathrm{Tr}(\varrho \log \varrho)$ denotes the von Neumann entropy, and $\varrho_{1} = \mathrm{Tr}_2(\varrho_{12})$ is the reduced density matrix of $M_1$ (analogously for $M_2$). In Fig.~\ref{fig_ex:MutualInfo}(a), we show simulation results for $\mathcal I(\varrho_{12})$ in the steady state, plotted as a function of the coupling strength $V$ and the frequency detuning between the two oscillators $\Delta\omega = \delta_2 - \delta_1$\,(see also Eq.~\ref{eq:EffectiveModel}). The simulation clearly shows the Arnold tongue feature \cite{pikovsky1985universal} that characterizes the synchronized regime, and further shows a threshold in the coupling strength $V$ below which synchronization is absent. Classically, the transition between synchronized and unsynchronized dynamics is a sharp boundary. In the presence of quantum fluctuations, however, the border of the Arnold tongue is smoothed~\cite{lee2013quantum, walter2014quantum}. Nevertheless, the mutual-information–based measure of synchronization remains a reliable indicator~\cite{ameriMutualInformationOrder2015a}.

\begin{figure*}[hbt!]
    \centering
    \includegraphics[width=\linewidth]{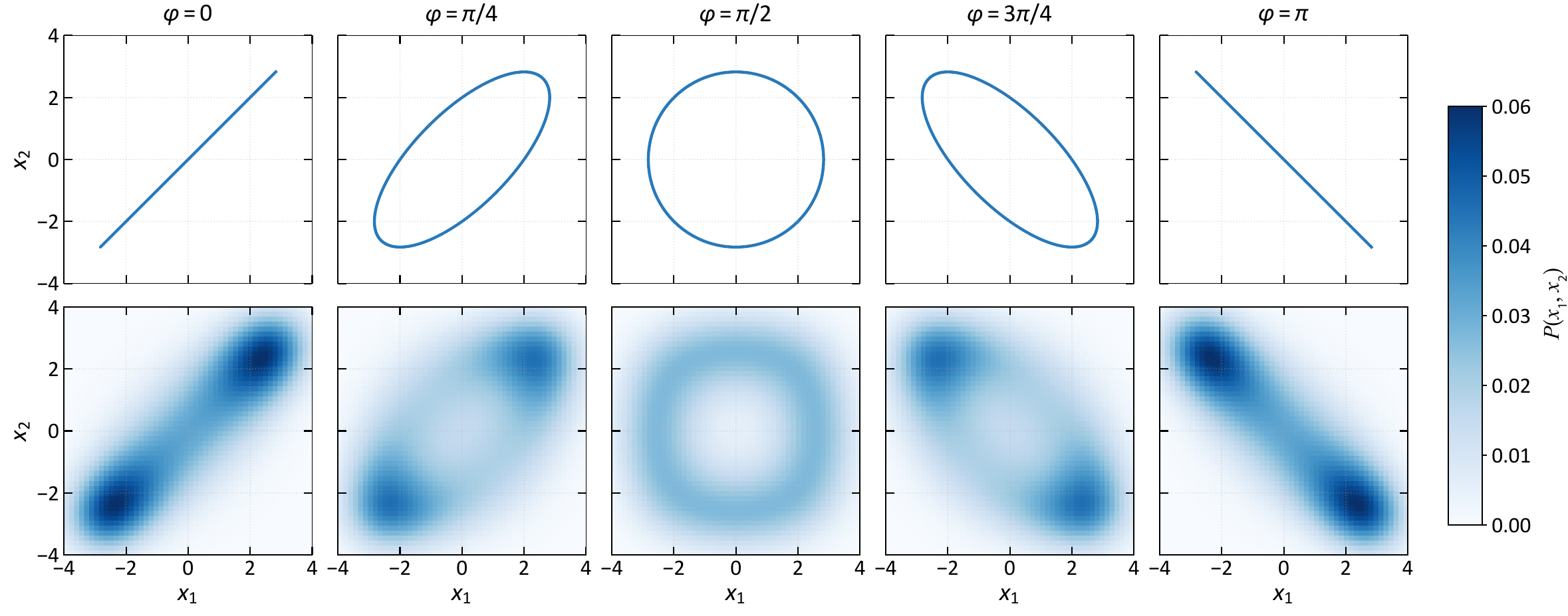}
    \caption{Classical trajectory in $x_1$--$x_2$ space (top row) and the joint probability distribution $P(x_1, x_2)$ between two oscillators (bottom row).}
    \label{fig:Lissajou}
\end{figure*}

However, evaluating $\mathcal I(\varrho_{12})$ requires access to the full density matrix $\varrho_{12}$. A more experimentally accessible quantity that captures similar information can be derived from the joint probability distributions between the quadratures. In particular, the mutual information of the joint $x_1$ and $x_2$ quadratures  given by
\begin{equation}
I(x_1:x_2) = \iint P(x_1, x_2) \log \left(\frac{P(x_1,x_2)}{P(x_1)P(x_2)}\right)~dx_1 dx_2,
\end{equation}
where $P(x_1, x_2) = \braket{x_1,x_2|\varrho_{12}|x_1,x_2}$ and $P(x_i)$ denotes the marginal distribution for $i=1,2$. Similarly, one can calculate the mutual information $I(x_1:p_2)$ from the joint distribution $P(x_1,p_2)$. The full mutual information would involve all quadratures, $x$ and $p$, simultaneously, i.e. $P(x_1, p_1, x_2, p_2)$, which is experimentally quite demanding. However, the combination $\mathcal{I} =I(x_1:x_2) + I(x_1:p_2)$ (sufficient due to symmetry considerations) provides the necessary information to identify features analogous to the classical Arnold tongue. In Fig.~\ref{fig_ex:MutualInfo}(b), we show simulation results for $\mathcal{I}$, which clearly exhibits the same overall structure as $\mathcal I(\varrho_{12})$. It thus serves as a sufficient indicator of synchronization in our setup. Fig.~\ref{fig_ex:MutualInfo}(c) shows a cut at $V/V_0 = 1.0$ as the frequency difference between the two oscillators increases, which demonstrates that $\mathcal{I}$ follows the same trend as $\mathcal{I}(\varrho_{12})$ and thus signals the presence of an Arnold tongue.\\

For the trivial fixed point $\alpha_1 = \alpha_2 = 0$, the eigenvalues are $\lambda_1 = \kappa_+$ and $\lambda_2 = \kappa_+ - 2V$. Thus, for any positive $\kappa_+$, at least one eigenvalue is positive, making this fixed point unstable. On the contrary, for the second fixed point, the eigenvalues become $\lambda_1 = -\kappa_+$ and $\lambda_2 = -\kappa_+ - 2V$, both of which are negative. Hence, this fixed point is stable, and nearby trajectories are attracted to it and eventually settle there in the long-time limit. In the original non-rotating frame, this corresponds to a synchronized solution of the form $\alpha_1(t) = R e^{-\mathrm i (\omega t - \varphi)} = e^{\mathrm i \varphi} \alpha_2(t)$, where we have set $\theta = 0$ for simplicity. Thus, at the mean-field level, the oscillators are synchronized with the same amplitude and a fixed phase difference determined by $\varphi$. As a result, changing the phase $\varphi$ yields simple Lissajous figures in the $x_1$–$x_2$ plane, where $x_i = (\alpha_i + \alpha_i^\ast)/\sqrt{2}$. These correspond to the simplest geometric figures: since amplitude and frequency are equal and only the phase varies, we obtain a diagonal line for $\varphi = 0$, an ellipse for intermediate $\varphi$, and a full circle for $\varphi = \pi/2$. We show the classical trajectory and the corresponding quantum solution $P(x_1,x_2)$ for different $\varphi$ in Fig.~\ref{fig:Lissajou}. Additionally, we compare these mean-field predictions with simulation results of the effective quantum model. Due to quantum noise, the numerical results exhibit a distribution rather than a single trajectory. However, the structure predicted by the classical mean-field model remains clearly visible.

\subsection{Classical and quantum dynamics for different synchronization phases.} 
In this section, we investigate how different synchronization phases affect the dynamics in the $x_1$--$x_2$ plane of Figs.\,2 and \ref{fig_SI_Px1x2}) by analyzing the mean-field equations. The classical dynamics corresponding to Eq.\,(1) is derived from the dynamics of $\alpha_i = \braket{a_i}$ by factorizing all expectation values (mean field). For simplicity, we consider equal oscillator frequencies, and the classical dynamics are given by the following equations 

\begin{equation}
\label{eq:CoupliedMF}
\begin{aligned}
\dot \alpha_1 &= \frac{\kappa_+}{2} \alpha_1 - \kappa_- \left|\alpha_1\right|^2\alpha_1 -\frac{V}{2}\left(\alpha_1 - e^{\mathrm i \varphi}\alpha_2\right),\\
\dot \alpha_2 &= \frac{\kappa_+}{2} \alpha_2 - \kappa_- \left|\alpha_2\right|^2\alpha_2 -\frac{V}{2}\left(\alpha_2 - e^{-\mathrm i \varphi}\alpha_1\right).
\end{aligned}
\end{equation}

The nonlinear system can be analyzed using linear stability analysis. To this end, we first identify the fixed points of Eq.~\ref{eq:CoupliedMF} and then analyze their stability by evaluating the Jacobian at those points. There are two sets of fixed points. The first is the trivial fixed point $\alpha_1 = \alpha_2 = 0$. The second set is given by $\alpha_1 = e^{\mathrm i \varphi} \alpha_2 = e^{\mathrm i (\varphi + \theta)} |\alpha|$, with positive amplitude $|\alpha| = \sqrt{\kappa_+ / (2 \kappa_-)}$ and $\theta$ a free global phase. The Jacobian $\mathbf J (\bar{\mathbf \alpha})$ evaluated at a fixed point $\bar{\mathbf \alpha}$=($\bar{\mathbf \alpha}_1$, $\bar{\mathbf \alpha}_2$) is given by

\begin{equation}
\mathbf J (\bar{\mathbf \alpha}) = \left(\begin{array}{cc}
\frac{\kappa_+}{2}-\kappa_- \left|\bar\alpha_1\right|^2 - \frac{V}{2} & e^{\mathrm i \varphi} V \\
e^{-\mathrm i \varphi} V & \frac{\kappa_+}{2}-\kappa_- \left|\bar\alpha_2\right|^2 - \frac{V}{2}
\end{array}
\right),
\end{equation}

which is Hermitian and all the eigenvalues are real. A fixed point is linearly stable if all eigenvalues of the Jacobian evaluated at that point are negative. 

For the trivial fixed point $\alpha_1 = \alpha_2 = 0$, the eigenvalues are $\lambda_1 = \kappa_+$ and $\lambda_2 = \kappa_+ - 2V$. Thus, for any positive $\kappa_+$, at least one eigenvalue is positive, making this fixed point unstable. On the contrary, for the second fixed point, the eigenvalues become $\lambda_1 = -\kappa_+$ and $\lambda_2 = -\kappa_+ - 2V$, both of which are negative. Hence, this fixed point is stable, and nearby trajectories are attracted to it and eventually settle there in the long-time limit. In the original non-rotating frame, this corresponds to a synchronized solution of the form $\alpha_1(t) = R e^{-\mathrm i (\omega t - \varphi)} = e^{\mathrm i \varphi} \alpha_2(t)$, where we have set $\theta = 0$ for simplicity. Thus, at the mean-field level, the oscillators are synchronized with the same amplitude and a fixed phase difference determined by $\varphi$. As a result, changing the phase $\varphi$ yields simple Lissajous figures in the $x_1$–$x_2$ plane, where $x_i = (\alpha_i + \alpha_i^\ast)/\sqrt{2}$. These correspond to the simplest geometric figures: since amplitude and frequency are equal and only the phase varies, we obtain a diagonal line for $\varphi = 0$, an ellipse for intermediate $\varphi$, and a full circle for $\varphi = \pi/2$. We show the classical trajectory and the corresponding quantum solution $P(x_1,x_2)$ for different $\varphi$ in Fig.~\ref{fig:Lissajou}. Additionally, we compare these mean-field predictions with simulation results of the effective quantum model. Due to quantum noise, the numerical results exhibit a distribution rather than a single trajectory. However, the structure predicted by the classical mean-field model remains clearly visible.

\bibliography{qsync_vdp}

\end{document}